\documentclass[aps,twocolumn,superscriptaddress,floatfix,prl]{revtex4-1}

\usepackage{lipsum}
\usepackage{graphicx}
\usepackage{amsmath,amssymb}
\usepackage{braket}
\usepackage{mathtools}
\usepackage{hyperref}
\usepackage{multirow}
\usepackage{color}
\usepackage[normalem]{ulem}
\usepackage{amsfonts}
\usepackage{float}
\usepackage{pdfpages} 
\makeatletter
\usepackage{subfigure}

\def\beq{\begin{equation}}
\def\eeq{\end{equation}}
\def\bea{\begin{eqnarray}}
\def\eea{\end{eqnarray}}

\AtBeginDocument{\let\LS@rot\@undefined}
\makeatother

\begin{document}

\title{Slow crossover from superdiffusion to diffusion in isotropic spin chains}

 \author{Catherine McCarthy}
\affiliation{Department of Physics, University of Massachusetts, Amherst, MA 01003, USA}

\author{Sarang Gopalakrishnan}
\affiliation{Department of Electrical and Computer Engineering, Princeton University, Princeton, NJ 08544, USA}

\author{Romain Vasseur}
\affiliation{Department of Physics, University of Massachusetts, Amherst, MA 01003, USA}

\begin{abstract}

Finite-temperature spin transport in integrable isotropic spin chains  (i.e., spin chains with continuous nonabelian symmetries) is known to be superdiffusive, with anomalous transport properties displaying remarkable robustness to isotropic integrability-breaking perturbations. Using a discrete-time classical model, we numerically study the crossover to conventional diffusion resulting from both noisy and Floquet isotropic perturbations of strength $\lambda$. We identify an anomalously-long crossover time scale $t_\star \sim \lambda^{-\alpha}$ with $\alpha \approx 6$ in both cases. We discuss our results in terms of a kinetic theory of transport that characterizes the lifetimes of large solitons responsible for superdiffusion. 

\end{abstract}
\vspace{1cm}

\maketitle

\paragraph{{\bf Introduction ---}} 

High-temperature transport in quantum magnets is usually assumed to be incoherent and diffusive. In this context, the discovery of superdiffusive spin transport with dynamical exponent $z=3/2$ in isotropic integrable quantum spin chains has been especially surprising~\cite{PhysRevLett.106.220601, lzp, Ljubotina2019, PhysRevLett.122.127202, gvw, Jepsen2020, Wei2022}.  While the full picture of superdiffusion in these systems is still coming into focus, a rapidly-growing body of work has elucidated several of its features ( see~\cite{Bertini_2021, Bulchandani2021, Gopalakrishnan_2023,gopalakrishnan2023superdiffusion} for recent reviews). Using recent developments from the theory of Generalized Hydrodynamics (GHD)~\cite{Fagotti, Doyon, Doyon_2020,Bastianello_2022,doyon2023generalized}, spin superdiffusion has been argued to emerge from the dynamics of large, semi-classical solitons of Goldstone-like nature~\cite{PhysRevLett.122.127202,vir2019, PhysRevLett.125.070601} that are generic in integrable models with continuous non-Abelian symmetries~\cite{PhysRevX.11.031023}. 
Numerical studies of these models show that the dynamical spin structure factor follows the Kardar-Parisi-Zhang~\cite{kpz} scaling form to high accuracy~\cite{PhysRevLett.122.210602,  10.21468/SciPostPhys.9.3.038, PhysRevB.102.115121, dupont_moore, Scheie2021, Wei2022}. However, describing higher-order fluctuations seems to require a more complicated fluctuating hydrodynamic theory~\cite{PhysRevLett.132.017101, Krajnik_2022,PhysRevLett.131.197102,rosenberg2023dynamics}.

 Superdiffusive spin transport was also observed in recent inelastic neutron scattering~\cite{Scheie2021}, cold atom quantum microscopy~\cite{Wei2022}, and superconducting qubit~\cite{rosenberg2023dynamics} experiments. Given that experiments are never described by perfectly integrable systems, it is natural to ask why superdiffusion appears to be so robust to integrability-breaking perturbations. In nearly integrable systems, the short-time dynamics are integrable, but at sufficiently long times, the dynamics become chaotic with diffusive transport properties~\cite{friedman2019diffusive,PhysRevLett.127.130601,bastianello2020generalised,Bastianello_2021}. The nature of this crossover from superdiffusive to diffusive spin transport in nearly-integrable isotropic spin chains remains poorly understood. Perturbative arguments indicate that the symmetry of the integrability-breaking perturbation is crucial: for anisotropic perturbations (those that break spin-rotation symmetry), the crossover timescales follow generic Fermi's Golden Rule (FGR) predictions and scale with the square of the perturbation strength~\cite{PhysRevLett_stability}.  In contrast, the timescales that characterize the crossover to diffusion associated with applying isotropic perturbations are known to be much slower~\cite{PhysRevLett_stability, roy2023nonequilibrium, PhysRevB.107.L100413, PhysRevB.105.L100403}, but these timescales have yet to be precisely characterized. 
This crossover is so slow that it remains controversial whether transport in the long-time limit is indeed diffusive
~\cite{dmki,10.21468/SciPostPhys.10.1.015,PhysRevLett.128.246603}. The primary obstacles to numerically observing the crossover to diffusion for isotropic perturbations have been the computational expense of simulating quantum systems and the robustness of superdiffusion with respect to isotropic perturbations, which together make accessing the crossover timescale difficult \cite{PhysRevLett_stability, PhysRevB.108.L081115}.  Even for classical spin chains simulated using standard numerical techniques like Runge-Kutta methods, superdiffusive transport in models like the Ishimori spin chain~\cite{Ishimori1982} survives for all numerically-accessible timescales in the presence of isotropic perturbations~\cite{roy2023nonequilibrium, PhysRevB.107.L100413, PhysRevB.105.L100403}.

In this letter, we employ an integrability-preserving discrete time integration scheme~\cite{Krajnik2020rot, 10.21468/SciPostPhys.9.3.038} that allows us to reach times up to $t_{\rm max}=2^{16}$ for systems of $N=200000$ spins, which is sufficient to observe the long timescale associated with the superdiffusive-to-diffusive crossover for isotropic perturbations.  For both noisy and Floquet isotropic perturbations of strength $\lambda$, we observe a crossover to conventional diffusion $t_\star \sim \lambda^{-\alpha}$ that is compatible with $\alpha=6$.  For noisy chains, this slow timescale can be understood in terms of a kinetic theory that characterizes the lifetimes of the large solitons responsible for spin transport.

\paragraph{{\bf Isotropic spin chains ---}} For concreteness we focus on integrable spin chains with Hamiltonian $H_0$, which are invariant under spin-rotation symmetry, perturbed by some spin-rotation-invariant integrability-breaking term $V$ of strength $\lambda$: 
\begin{equation} \label{eqHeisenberg}
H = (1 - \lambda) H_0 + \lambda V.
\end{equation}
Here $\lambda \in [0,1]$ is a parameter interpolating between integrable ($\lambda=0$) and purely non-integrable ($\lambda=1$) dynamics. Our analysis is general for any integrable spin chain, classical or quantum, that is invariant under some continuous nonabelian symmetry. However, since diffusion emerges on very long timescales, we simulate models for which one can reliably study very late times. Thus, first, we consider classical spin chains: Each site $n$ along the classical spin chain hosts an $O(3)$  vector of unit norm $\vec{S}_n = (S^x_n, S^y_n, S^z_n)$ with canonical Poisson brackets $\lbrace S^i_n, S^j_m \rbrace = \epsilon_{ijk} S^k_n \delta_{nm}$, subject to nearest-neighbor ferromagnetic interactions. Second, to avoid the accumulation of truncation errors in schemes like Runge-Kutta, we will simulate a discrete-time, Trotterized form of the dynamics. To this end we will exploit the existence of Trotterizations of classical spin chains that preserve integrability~\cite{Krajnik2020rot, 10.21468/SciPostPhys.9.3.038, Krajnik_2022}.

In the remainder of this work, the integrable dynamics will be generated by an integrable Trotterization~\cite{Krajnik2020rot, 10.21468/SciPostPhys.9.3.038, Krajnik_2022} of the classical Ishimori Hamiltonian  $H_0 = -\sum_{n} \mathrm{ln}( 1 + \vec{S}_n \cdot \vec{S}_{n+1})$~\cite{Ishimori1982}. The isotropic integrability-breaking perturbation will be an (again Trotterized) Heisenberg interaction $V= -\sum_{n} g_{n}(t) \vec{S}_n \cdot \vec{S}_{n+1}$. We will consider two cases: (i)~perturbations that are noisy, where $g_{n}(t)$ will be taken to be random variables uncorrelated in both space and time, $\overline{g_{n}(t) g_{m}(t') } = \delta_{nm} \delta(t-t')$, where $\overline{O}$ denotes the average of $O$ over noise, and (ii)~perturbations that are time-periodic, for which $g_n(t) = 1$.

\paragraph{{\bf Spin transport ---}} Numerical, theoretical and experimental studies have shown that spin transport in the integrable limit $\lambda=0$ is known to be {\em superdiffusive}, while energy transport is purely ballistic \cite{PhysRevLett.106.220601, lzp, Ljubotina2019, PhysRevLett.122.127202, gvw, Scheie2021, Jepsen2020, Wei2022}. For simplicity, we will focus on the infinite-temperature regime where spins are initialized at random, although our conclusions carry over to arbitrary finite temperatures. In order to characterize spin transport, we will focus on fluctuations of spin transfer across a given bond in a background equilibrium state. We define spin transfer as the time-integrated spin current across the link between sites $n$ and $n+1$, $Q(t) = \int_0^t dt j_n(t)$, where $j_n(t)$ is the spin current between sites $n$ and $n+1$, defined by the continuity equation $\partial_t S^z_n + j_{n+1} - j_n = 0$. We invoke translation invariance and average this quantity over each bond $n$ to improve statistics.  Since there is no net spin transfer in a background equilibrium state, the thermal average $\langle Q(t) \rangle = 0$ vanishes.  Fluctuations in spin transfer can be used to characterize spin transport through the relation
\begin{equation} \label{eqChargetransfer}
\langle Q(t)^2 \rangle \sim t^{1/z},
\end{equation}
where $z$ is the {\em dynamical exponent}, which characterizes transport by relating space and time scaling through $t \sim x^{z}$. For diffusive transport, we see the dynamical exponent $z=2$, whereas we see $z=3/2$ in integrable isotropic chains, corresponding to {\em superdiffusive} (faster than diffusive) transport with an effective time-dependent diffusion constant $D(t) \sim t^{1/3}$. In order to characterize the crossover from $z=3/2$ to $z=2$ upon applying an integrability-breaking perturbation, we will define the time-dependent dynamical exponent as the logarithmic derivative  $z(t) = \left(\frac{d \ln \langle Q^2 \rangle }{d \ln t}\right)^{-1}$. 

As a consistency check, we also characterize spin transport and the crossover to diffusion using the spin autocorrelation function $C(t) = \langle S^z_n(t) S^z_n(0)  \rangle \sim t^{-1/z}$.   We extract the dynamical exponent $z$ from $C(t)$ and observe results compatible with those discussed below (see \cite{suppmat}).  Since transport becomes diffusive at long times with $C(t) \sim 1/\sqrt{D(\lambda) t}$, the autocorrelation function can also be used to compute the diffusion constant $D(\lambda)$ from $\lim_{t \to \infty} \left( \sqrt{t}~ C(t) \right)$ \cite{10.21468/SciPostPhys.10.1.015}.

\begin{figure*}[t!]
	\centering 
	\includegraphics[width=.99\linewidth]{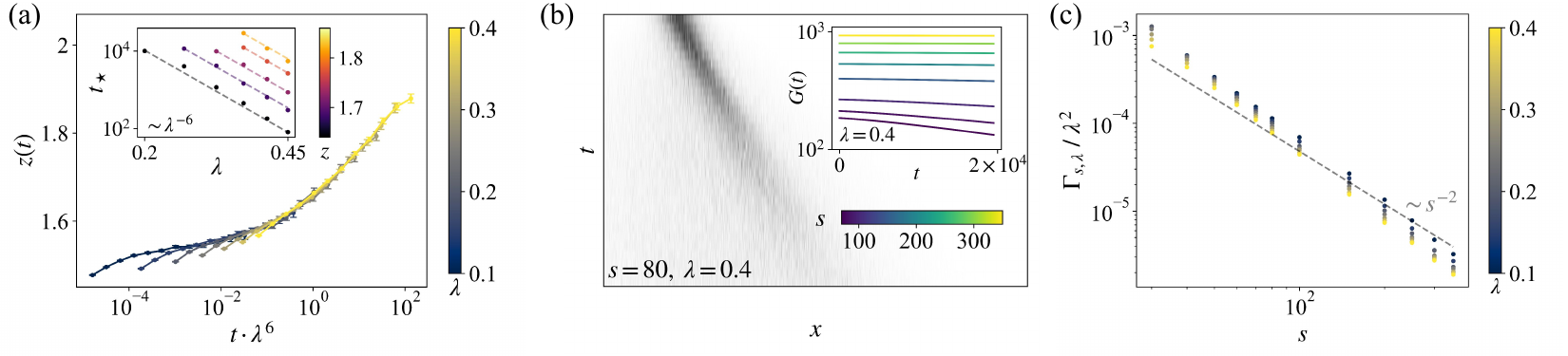}
 
	\caption{{\bf Crossover to diffusion in noisy spin chains.} (a) Effective time-dependent dynamical exponent $z(t)$ extracted from spin transfer fluctuations, showing a crossover from $z=3/2$ (superdiffusion, integrable case) to $z=2$ (diffusion) for different values of the integrability-breaking parameter $\lambda$. Data is averaged over at least 2000 samples (including random initial states and different realizations of the noisy perturbation), and we used a timestep of $\tau=0.25$. We observe a collapse as a function of the rescaled variable $t \lambda^6$, indicating a crossover time scale $t_\star \sim \lambda^{-6}$. Inset: time scale $t_\star(\lambda)$ at which $z(t)$ reaches some set value of $z$ (color scale), showing a clear power-law scaling $t_\star \sim \lambda^{-6}$. (b) Example of soliton ``melting'' in the presence of a noisy perturbation. Inset: Exponential decay of the IPR $G(t) $ for different soliton sizes $s$. (c) Soliton decay rates in the vacuum, showing a scaling compatible with $\Gamma_{s,\lambda} \sim \lambda^2/s^2$. }
	\label{fig:noisy}
\end{figure*}

\paragraph{{\bf Noisy perturbations: numerics ---}}
The discrete-time model used here can be efficiently simulated up to very long times without accumulating Trotter errors, allowing us to reach maximum times of $t=2^{16}$ for systems up to $N=200000$ spins, which is sufficient to observe the crossover to diffusion for noisy isotropic perturbations.  The effective time-dependent dynamical exponent $z(t)$ extracted from the spin transfer shows a clear crossover from $z=3/2$ to $z=2$ at long times (Fig.~\ref{fig:noisy}a). We find that the crossover occurs over a  long time scale $t_\star \sim \lambda^{-\alpha}$ with an exponent consistent with $\alpha = 6$. We additionally extract the diffusion constant $D(\lambda)$ from the autocorrelation function $C(t) = \frac{1}{N} \sum_n \langle S_n^z(t) S_n^z(0) \rangle$, and find that it scales as $D(\lambda) \sim \lambda^{-2}$ (see \cite{suppmat}).  Since matching the diffusive and integrable behaviors at $t=t_\star$ gives $D(\lambda) \sim t_\star^{1/3}\sim \lambda^{-\alpha/3}$  for $t_\star \sim \alpha$, the scaling $D(\lambda) \sim \lambda^{-2}$ is also consistent with $t_\star \sim \lambda^{-6}$ \cite{suppmat}. (In contrast, a process by which the solitons decay in an $s$-independent way through Golden Rule processes yields a crossover time $t_\star \sim \lambda^{-2}$ and a diffusion constant $D(\lambda) \sim \lambda^{-2/3}$, consistent with numerical data on \emph{anisotropic} perturbations~\cite{PhysRevLett_stability, suppmat}.)


\paragraph{{\bf Kinetic theory ---}} To understand this unusually long crossover time scale, we turn to a kinetic theory of the quasiparticles responsible for superdiffusion in the integrable case. In integrable isotropic spin chains, superdiffusive transport with $z=3/2$ is well established in terms of the system's quasiparticle excitations~\cite{PhysRevLett.122.127202,gvw,PhysRevLett.123.186601, PhysRevLett.123.186601, PhysRevLett.125.070601, PhysRevX.11.031023}. In classical chains, these quasiparticles are infinitely long-lived solitons which remain stable even at high temperatures. Solitons are characterized by their size $s$, a parameter that is quantized in quantum spin chains and continuous in classical chains. These solitons have velocity $v_s \sim 1/s$, so large solitons move slowly, and occur with density $\rho_s \sim 1/s^3$ in thermal equilibrium \cite{idmp}. They also carry spin $m_s =s$ in the vacuum, although this net magnetization can be screened in thermal states by a background consisting of other overlapping solitons \cite{PhysRevLett.78.943, PhysRevB.57.8307, idmp}. Integrability results show that solitons are screened at a rate $\Gamma^0_s \sim s^{-3}$ \cite{PhysRevLett.122.127202, gvw,PhysRevB.101.224415}. This means that at a given time $t$, small solitons with $s \ll s_\star(t) \sim t^{1/3}$ are fully screened and carry a net magnetization $m_s=0$, and so they do not contribute to spin transport.  On the other hand, ``giant'' solitons $s \gg s_\star(t) \sim t^{1/3}$ remain charged and dominate transport. Combined with the slow velocity $v_s \sim 1/s$ of giant solitons, this screening mechanism is responsible for the superdiffusion exponent $z=3/2$ in integrable isotropic chains \cite{PhysRevLett.122.127202}.

\paragraph{{\bf Soliton decay rates ---} }

In the presence of an integrability-breaking perturbation of strength $\lambda \neq 0$, solitons acquire a finite lifetime due to both vacuum decay processes and backscattering from collisions.  The finite lifetime of a soliton of size $s$ subjected to an integrability-breaking perturbation is characterized by the decay rate $\Gamma_{s,\lambda}$, which we expect to take the form $\Gamma_{s, \lambda} \sim \sum_n \lambda^{\alpha_n} s^{-\beta_n},$ with each term indexed by $n$ corresponding to a different decay process.  Note any process with $\beta_n \geq 3$ will occur slower than the integrable screening process $\Gamma^0_s \sim s^{-3}$ for large solitons and is thus irrelevant. The crossover to diffusion can be understood as a competition between the integrable screening rate $\Gamma^0_s \sim s^{-3}$ and the perturbation-induced decay rate $\Gamma_{s, \lambda}$.  The crossover occurs when perturbative contributions to a soliton's decay rate $\Gamma_{s, \lambda}$ overpower the screening rate $\Gamma^0_s$; that is, the time at which the perturbative decay rate scales like $\Gamma_{s, \lambda} \sim s^{-3}$.

The task of understanding the crossover therefore reduces to finding the leading-order term of the decay rate $\Gamma_{s, \lambda}$ induced by a noisy isotropic perturbation.  In general, analytically evaluating $\Gamma_{s,\lambda}$ is very complicated, even perturbatively. We study this question numerically by considering the decay of a single soliton of size $s$ in a vacuum background state, corresponding to an initial state $(S^x_n, S^y_n, S^z_n) = (\sqrt{1-(S_n^z)^2}\cos \phi_n, \sqrt{1-(S_n^z)^2} \sin \phi_n, S_n^z)$ with
\begin{equation} \label{eqSoliton}
      S_n^z = \tanh^2 \left( \frac{n}{2s} \right), 
\end{equation}
and $\phi_n = \arctan \left(  \tanh\frac{n}{2s} \right) + \frac{n}{2s}$ \cite{lrt}. This initial state corresponds to an exact right-moving soliton in the Ishimori chain with interaction $\mathrm{ln}( 1 + \vec{S}_n \cdot \vec{S}_{n+1}) $ -- left-movers can be obtained by shifting the phase $\phi_n \rightarrow \phi_n + \pi/2$.  In the context of our discrete time numerics, this initial state is not technically an exact soliton (even for $\lambda=0$), but it has a very strong overlap with the solitons of the discrete-time model. 

In the integrable case ($\lambda=0$), solitons propagate without decaying. Applying a noisy perturbation of strength $\lambda \neq 0$ causes a giant soliton to disintegrate into small solitons that diffuse through backscattering. To quantify this decay, we introduce the inverse participation ratio (IPR) $ G(t) = \sum_n (S_n^z(t) -1)^2$.  Spin conservation fixes  $\sum_n (S_n^z -1)$ as a time-independent constant of order $s$; therefore, $G(t)$ provides a metric to quantify the localization of soliton.  For a localized soliton, $G(t)$ remains a constant of order $s$, while for a delocalized soliton it decays with system size as $1/L$. For large systems and long times, we observe an exponential decay $G(t) \sim {\rm e}^{- \Gamma_{s,\lambda} t}$, which allows us to extract the decay rates $\Gamma_{s,\lambda}$ numerically for various soliton sizes $s$ and perturbation strengths $\lambda$ (Fig.~\ref{fig:noisy}(b)). We find that 
\begin{equation} \label{eq:ratevacuum}
\Gamma_{s,\lambda} \sim \frac{\lambda^2}{s^2},
\end{equation}
in agreement with recent perturbative arguments (Fig.~\ref{fig:noisy}(c)) \cite{PhysRevLett_stability}. The long lifetimes $\sim s^{2}$ of large solitons can be understood in terms of the Goldstone-like nature of the solitons~\eqref{eqSoliton}: matrix elements of isotropic perturbations are suppressed as $1/s$ acting on a soliton of size $s$ \cite{PhysRevLett_stability}. Comparing the decay rates~\eqref{eq:ratevacuum} to the integrable screening rates, we find the crossover soliton size $s_\star \sim \lambda^{-2}$, and the crossover time scale 
\begin{equation}
t_\star \sim \lambda^{-6},
\end{equation}
in agreement with the anomalously long time scale observed in our transport data (Fig.~\ref{fig:noisy}(a)) and extracted diffusion constant (see \cite{suppmat}).

While the vacuum decay rates~\eqref{eq:ratevacuum} fully explain the crossover time scale $t_\star \sim \lambda^{-6}$, they also lead to logarithmic corrections to diffusion, since the giant solitons' lifetime $\sim s^{2}$ in the vacuum is long enough to lead to anomalous behavior \cite{PhysRevLett_stability}. Whether interactions can induce decay rates $\sim s^{-\alpha}$ with $\alpha < 2$ that restore diffusion remains an important question.

\begin{figure}[t!]
    \includegraphics[width=0.99\columnwidth]{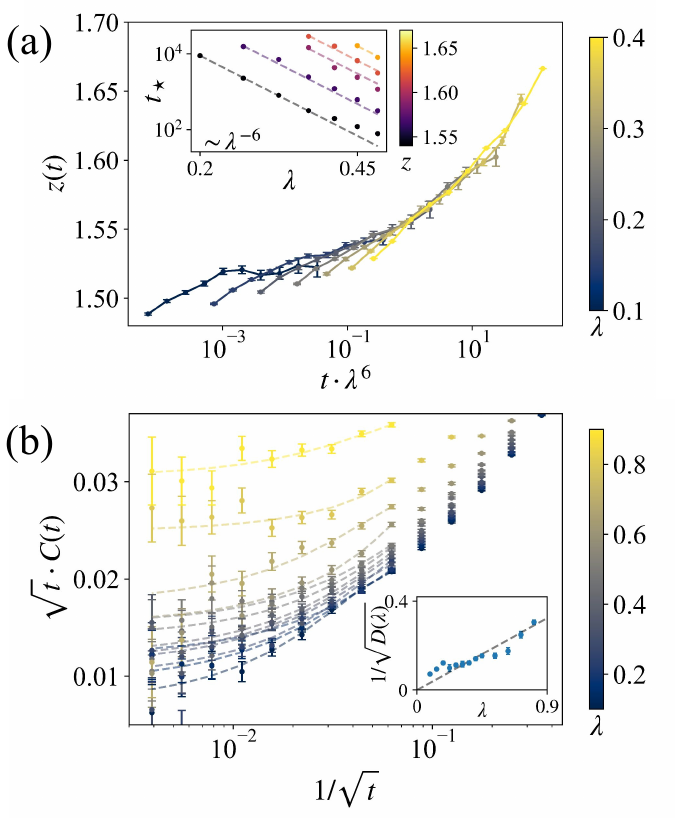}
    \caption{\textbf{Crossover to diffusion for Floquet
    perturbations.} (a) dynamical exponent $z(t)$ extracted from fluctuations of spin transfer plotted against $t\lambda^6$. Data is averaged over at least 4000 random initial states for a system size of $N=200000$ sites with a timestep $\tau=0.25$.  We observe a curve collapse is compatible with crossover timescale $t_\star \sim \lambda^{-6}.$  Inset: timescale $t_\star$ at which a particular value $z(t_\star)$ is reached for different lambdas, demonstrating power-law scaling $\sim \lambda^{-6}$. (b) Adjusted autocorrelation function $\sqrt{t}C(t) = \frac{\sqrt{t}}{N} \sum_n \langle S_n^z(t) S_n^z(0) \rangle$ plotted against $1/\sqrt{t}$.  Curves corresponding to large perturbations of strength $\lambda>0.5$ are averaged over at least 2000 random initial states, while small perturbations $\lambda \leq 0.5$ are averaged over at least 4000 random initial states.  Dashed lines correspond to fitting $\sqrt{t}C(t)$ for $t \geq 256$ with its standard form at late times, $\sqrt{t}C(t) \sim 1/\sqrt{D(\lambda)} + A/\sqrt{t} + B/t + \dots$, where $D(\lambda)$ is the diffusion constant up to irrelevant numerical prefactors~\cite{10.21468/SciPostPhys.10.1.015}.  Inset: $1/\sqrt{D(\lambda)}$ extracted by fitting the adjusted autocorrelation function $\sqrt{t} C(t)$.  The linear scaling of $1/\sqrt{D(\lambda)}$ is consistent with a crossover timescale of $t_\star \sim \lambda^{-6}$~\cite{suppmat}.}
    \label{fig:clean}

\end{figure}

\paragraph{{\bf Floquet perturbations --- }}

The previous section considered noisy perturbations $V(t) = -\sum_n g_n(t) \vec{S}_n \cdot \vec{S}_{n+1}$, with $g_n(t)$ uncorrelated random variables in both space and time. We now consider non-noisy (Floquet) perturbations with $g_n(t) = 1$.  The crossover associated with a noisy perturbation is primarily driven by the vacuum decay rate $\Gamma_{s, \lambda} \sim \lambda^2/s^2$; however, in the absence of noise, isolated solitons remain stable even in the non-integrable classical Heisenberg chain~\cite{PhysRevE.106.L062202}.  The largest contribution to the decay rate must therefore arise from many-body soliton scattering processes; this term should also exist in the decay rate for the noisy perturbations but is masked by the leading-order contribution from the vacuum decay process.  Therefore, we expect the crossover timescale associated with Floquet perturbations to be at least as slow as the noisy crossover timescale $t_\star \sim \lambda^{-6}$, if not slower.  This picture is consistent with the recent numerical works on classical spin chains \cite{roy2023nonequilibrium, PhysRevB.107.L100413, PhysRevB.105.L100403}, which were unable to observe a systematic crossover towards diffusion on timescales accessible by standard continuous-time integration schemes like adaptive Runge-Kutta methods.  

The discrete-time numerics used here are able to access very long times without finite time-step errors, which allows us to observe timescales inaccessible to Runge-Kutta setups.  Even so, we observe a curve collapse for the dynamical exponent $z(t)$ that is slower than the noisy case but still consistent with $t_\star \sim \lambda^{-\alpha}$ scaling with $\alpha \approx 6$ (Fig.~\ref{fig:clean}a). For consistency, we extract the diffusion constant $D(\lambda)$ from the autocorrelation function using the same method as described for the noisy perturbation, which is also compatible with $t_\star \sim \lambda^{-6}$ scaling (Fig. \ref{fig:clean}b).  We note that our data cannot definitively exclude a different exponent in the true scaling limit $t \to \infty $ and $\lambda \to 0$, and different values of $\alpha$ lead to acceptable collapses as well (see \cite{suppmat}).  

As previously discussed, we expect that the crossover to diffusion is driven by collisions between multiple solitons; however, since the IPR is not a meaningful metric for a finite density of solitons, it is difficult to determine collision decay rates using the same methods as were used to extract the noisy vacuum decay rate.  
By simulating experiments in which large solitons first interact with a non-trivial background under $\lambda \neq 0$ dynamics, and then filtering out the background by setting $\lambda=0$ and allowing it to ``expand'' into a vacuum region, we are able to show that the soliton lifetime increases with $s$, although the data are too noisy to extract a reliable exponent~\cite{suppmat}.
We note that the crossover scale $t_\star \sim \lambda^{-6}$ is compatible with higher-order perturbative decay rates $\Gamma_{s,\lambda} \sim \lambda^{2(n+1)} s^{n-2}$ for any $n$. On general grounds, we expect the collisions between $s$-sized solitons at large $s$ and $O(1)$-sized solitons to scale as $1/s^2$, since a small soliton experiences a large one as a local vacuum rotation. In effect, a thermal background of small solitons acts as noise on the large ones, relating the Floquet and noisy cases. This process would once again lead to logarithmically corrected diffusion. Any higher-order processes with $n > 0$, if present, would lead to a full crossover to diffusion. Determining the form of the decay rates that emerge from interactions remains an important challenge for future work.

{\bf Discussion ---}  We have numerically observed the slow crossover from superdiffusive to diffusive transport for isotropically-perturbed integrable classical spin chains, with a crossover timescale $t_\star \sim \lambda^{-\alpha}$ with $\alpha \approx 6$.  Our ability to reach this anomalously long crossover timescale is due to the classical discrete time algorithm used in this work, in contrast to previous work that studied this question with matrix-product states or classical continuous time numerical methods.  The $t_\star \sim \lambda^{-6}$ scaling holds for both noisy and Floquet perturbations.  While the exponent $\alpha=6$ is expected perturbatively in the noisy case, and has a heuristic justification in the clean case, 
a full explanation of the crossover to diffusion in either case 
requires analyzing the decay rates that arise from $n$-body soliton scattering processes, and presents an important challenge for future theoretical work. For example it is possible that the crossover to diffusion happens in multiple stages, with an intermediate regime of logarithmically corrected diffusion that gets parametrically large in $\lambda$. We note that although our numerics are consistent with an exponent $\alpha \approx 6$ for Floquet perturbations, we cannot conclusively exclude other values of $\alpha$ in the limit of $\lambda \rightarrow 0$ and hope that future theoretical and numerical advances will pinpoint the exact exponent.

\paragraph{\textbf { Acknowledgements -- }} We thank Jacopo De Nardis and Brayden Ware for collaborations on related topics, and Adam McRoberts and Roderich Moessner for helpful discussions. C.M. acknowledges support from NSF GRFP-1938059. This work was supported by NSF grants DMR-2103938 (S.G.) and DMR-2104141 (R.V.).

\paragraph {\textit{Note:}} During the completion of this work, we became aware of a related work by Adam McRoberts and Roderich Moessner~\cite{adams_draft} reporting a crossover timescale of $t_{\star} \sim \lambda^{-3}$ in an energy-conserving model and analyzing its temperature dependence.  Future work would be needed to understand why the crossovers in both models appear to be dominated by different decay processes on accessible time scales.

\bibliography{refs}

\begin{thebibliography}{52}%
\makeatletter
\providecommand \@ifxundefined [1]{%
 \@ifx{#1\undefined}
}%
\providecommand \@ifnum [1]{%
 \ifnum #1\expandafter \@firstoftwo
 \else \expandafter \@secondoftwo
 \fi
}%
\providecommand \@ifx [1]{%
 \ifx #1\expandafter \@firstoftwo
 \else \expandafter \@secondoftwo
 \fi
}%
\providecommand \natexlab [1]{#1}%
\providecommand \enquote  [1]{``#1''}%
\providecommand \bibnamefont  [1]{#1}%
\providecommand \bibfnamefont [1]{#1}%
\providecommand \citenamefont [1]{#1}%
\providecommand \href@noop [0]{\@secondoftwo}%
\providecommand \href [0]{\begingroup \@sanitize@url \@href}%
\providecommand \@href[1]{\@@startlink{#1}\@@href}%
\providecommand \@@href[1]{\endgroup#1\@@endlink}%
\providecommand \@sanitize@url [0]{\catcode `\\12\catcode `\$12\catcode `\&12\catcode `\#12\catcode `\^12\catcode `\_12\catcode `\%12\relax}%
\providecommand \@@startlink[1]{}%
\providecommand \@@endlink[0]{}%
\providecommand \url  [0]{\begingroup\@sanitize@url \@url }%
\providecommand \@url [1]{\endgroup\@href {#1}{\urlprefix }}%
\providecommand \urlprefix  [0]{URL }%
\providecommand \Eprint [0]{\href }%
\providecommand \doibase [0]{http://dx.doi.org/}%
\providecommand \selectlanguage [0]{\@gobble}%
\providecommand \bibinfo  [0]{\@secondoftwo}%
\providecommand \bibfield  [0]{\@secondoftwo}%
\providecommand \translation [1]{[#1]}%
\providecommand \BibitemOpen [0]{}%
\providecommand \bibitemStop [0]{}%
\providecommand \bibitemNoStop [0]{.\EOS\space}%
\providecommand \EOS [0]{\spacefactor3000\relax}%
\providecommand \BibitemShut  [1]{\csname bibitem#1\endcsname}%
\let\auto@bib@innerbib\@empty
\bibitem [{\citenamefont {\ifmmode \check{Z}\else \v{Z}\fi{}nidari\ifmmode~\check{c}\else \v{c}\fi{}}(2011)}]{PhysRevLett.106.220601}%
  \BibitemOpen
  \bibfield  {author} {\bibinfo {author} {\bibfnamefont {M.}~\bibnamefont {\ifmmode \check{Z}\else \v{Z}\fi{}nidari\ifmmode~\check{c}\else \v{c}\fi{}}},\ }\href {\doibase 10.1103/PhysRevLett.106.220601} {\bibfield  {journal} {\bibinfo  {journal} {Phys. Rev. Lett.}\ }\textbf {\bibinfo {volume} {106}},\ \bibinfo {pages} {220601} (\bibinfo {year} {2011})}\BibitemShut {NoStop}%
\bibitem [{\citenamefont {Ljubotina}\ \emph {et~al.}(2017)\citenamefont {Ljubotina}, \citenamefont {{\v Z}nidari{\v c}},\ and\ \citenamefont {Prosen}}]{lzp}%
  \BibitemOpen
  \bibfield  {author} {\bibinfo {author} {\bibfnamefont {M.}~\bibnamefont {Ljubotina}}, \bibinfo {author} {\bibfnamefont {M.}~\bibnamefont {{\v Z}nidari{\v c}}}, \ and\ \bibinfo {author} {\bibfnamefont {T.}~\bibnamefont {Prosen}},\ }\href {http://dx.doi.org/10.1038/ncomms16117} {\bibfield  {journal} {\bibinfo  {journal} {Nature Communications}\ }\textbf {\bibinfo {volume} {8}},\ \bibinfo {pages} {16117 EP } (\bibinfo {year} {2017})}\BibitemShut {NoStop}%
\bibitem [{\citenamefont {Ljubotina}\ \emph {et~al.}(2019{\natexlab{a}})\citenamefont {Ljubotina}, \citenamefont {Zadnik},\ and\ \citenamefont {Prosen}}]{Ljubotina2019}%
  \BibitemOpen
  \bibfield  {author} {\bibinfo {author} {\bibfnamefont {M.}~\bibnamefont {Ljubotina}}, \bibinfo {author} {\bibfnamefont {L.}~\bibnamefont {Zadnik}}, \ and\ \bibinfo {author} {\bibfnamefont {T.}~\bibnamefont {Prosen}},\ }\href {\doibase 10.1103/physrevlett.122.150605} {\bibfield  {journal} {\bibinfo  {journal} {Physical Review Letters}\ }\textbf {\bibinfo {volume} {122}} (\bibinfo {year} {2019}{\natexlab{a}}),\ 10.1103/physrevlett.122.150605}\BibitemShut {NoStop}%
\bibitem [{\citenamefont {Gopalakrishnan}\ and\ \citenamefont {Vasseur}(2019)}]{PhysRevLett.122.127202}%
  \BibitemOpen
  \bibfield  {author} {\bibinfo {author} {\bibfnamefont {S.}~\bibnamefont {Gopalakrishnan}}\ and\ \bibinfo {author} {\bibfnamefont {R.}~\bibnamefont {Vasseur}},\ }\href {\doibase 10.1103/PhysRevLett.122.127202} {\bibfield  {journal} {\bibinfo  {journal} {Phys. Rev. Lett.}\ }\textbf {\bibinfo {volume} {122}},\ \bibinfo {pages} {127202} (\bibinfo {year} {2019})}\BibitemShut {NoStop}%
\bibitem [{\citenamefont {Gopalakrishnan}\ \emph {et~al.}(2019)\citenamefont {Gopalakrishnan}, \citenamefont {Vasseur},\ and\ \citenamefont {Ware}}]{gvw}%
  \BibitemOpen
  \bibfield  {author} {\bibinfo {author} {\bibfnamefont {S.}~\bibnamefont {Gopalakrishnan}}, \bibinfo {author} {\bibfnamefont {R.}~\bibnamefont {Vasseur}}, \ and\ \bibinfo {author} {\bibfnamefont {B.}~\bibnamefont {Ware}},\ }\href {\doibase 10.1073/pnas.1906914116} {\bibfield  {journal} {\bibinfo  {journal} {Proceedings of the National Academy of Sciences}\ }\textbf {\bibinfo {volume} {116}},\ \bibinfo {pages} {16250} (\bibinfo {year} {2019})}\BibitemShut {NoStop}%
\bibitem [{\citenamefont {Jepsen}\ \emph {et~al.}(2020)\citenamefont {Jepsen}, \citenamefont {Amato-Grill}, \citenamefont {Dimitrova}, \citenamefont {Ho}, \citenamefont {Demler},\ and\ \citenamefont {Ketterle}}]{Jepsen2020}%
  \BibitemOpen
  \bibfield  {author} {\bibinfo {author} {\bibfnamefont {P.~N.}\ \bibnamefont {Jepsen}}, \bibinfo {author} {\bibfnamefont {J.}~\bibnamefont {Amato-Grill}}, \bibinfo {author} {\bibfnamefont {I.}~\bibnamefont {Dimitrova}}, \bibinfo {author} {\bibfnamefont {W.~W.}\ \bibnamefont {Ho}}, \bibinfo {author} {\bibfnamefont {E.}~\bibnamefont {Demler}}, \ and\ \bibinfo {author} {\bibfnamefont {W.}~\bibnamefont {Ketterle}},\ }\href {\doibase 10.1038/s41586-020-3033-y} {\bibfield  {journal} {\bibinfo  {journal} {Nature}\ }\textbf {\bibinfo {volume} {588}},\ \bibinfo {pages} {403} (\bibinfo {year} {2020})}\BibitemShut {NoStop}%
\bibitem [{\citenamefont {Wei}\ \emph {et~al.}(2022)\citenamefont {Wei}, \citenamefont {Rubio-Abadal}, \citenamefont {Ye}, \citenamefont {Machado}, \citenamefont {Kemp}, \citenamefont {Srakaew}, \citenamefont {Hollerith}, \citenamefont {Rui}, \citenamefont {Gopalakrishnan}, \citenamefont {Yao}, \citenamefont {Bloch},\ and\ \citenamefont {Zeiher}}]{Wei2022}%
  \BibitemOpen
  \bibfield  {author} {\bibinfo {author} {\bibfnamefont {D.}~\bibnamefont {Wei}}, \bibinfo {author} {\bibfnamefont {A.}~\bibnamefont {Rubio-Abadal}}, \bibinfo {author} {\bibfnamefont {B.}~\bibnamefont {Ye}}, \bibinfo {author} {\bibfnamefont {F.}~\bibnamefont {Machado}}, \bibinfo {author} {\bibfnamefont {J.}~\bibnamefont {Kemp}}, \bibinfo {author} {\bibfnamefont {K.}~\bibnamefont {Srakaew}}, \bibinfo {author} {\bibfnamefont {S.}~\bibnamefont {Hollerith}}, \bibinfo {author} {\bibfnamefont {J.}~\bibnamefont {Rui}}, \bibinfo {author} {\bibfnamefont {S.}~\bibnamefont {Gopalakrishnan}}, \bibinfo {author} {\bibfnamefont {N.~Y.}\ \bibnamefont {Yao}}, \bibinfo {author} {\bibfnamefont {I.}~\bibnamefont {Bloch}}, \ and\ \bibinfo {author} {\bibfnamefont {J.}~\bibnamefont {Zeiher}},\ }\href {\doibase 10.1126/science.abk2397} {\bibfield  {journal} {\bibinfo  {journal} {Science}\ }\textbf {\bibinfo {volume} {376}},\ \bibinfo {pages} {716} (\bibinfo {year} {2022})}\BibitemShut {NoStop}%
\bibitem [{\citenamefont {Bertini}\ \emph {et~al.}(2021)\citenamefont {Bertini}, \citenamefont {Heidrich-Meisner}, \citenamefont {Karrasch}, \citenamefont {Prosen}, \citenamefont {Steinigeweg},\ and\ \citenamefont {\ifmmode \check{Z}\else \v{Z}\fi{}nidari\ifmmode~\check{c}\else \v{c}\fi{}}}]{Bertini_2021}%
  \BibitemOpen
  \bibfield  {author} {\bibinfo {author} {\bibfnamefont {B.}~\bibnamefont {Bertini}}, \bibinfo {author} {\bibfnamefont {F.}~\bibnamefont {Heidrich-Meisner}}, \bibinfo {author} {\bibfnamefont {C.}~\bibnamefont {Karrasch}}, \bibinfo {author} {\bibfnamefont {T.}~\bibnamefont {Prosen}}, \bibinfo {author} {\bibfnamefont {R.}~\bibnamefont {Steinigeweg}}, \ and\ \bibinfo {author} {\bibfnamefont {M.}~\bibnamefont {\ifmmode \check{Z}\else \v{Z}\fi{}nidari\ifmmode~\check{c}\else \v{c}\fi{}}},\ }\href {\doibase 10.1103/RevModPhys.93.025003} {\bibfield  {journal} {\bibinfo  {journal} {Rev. Mod. Phys.}\ }\textbf {\bibinfo {volume} {93}},\ \bibinfo {pages} {025003} (\bibinfo {year} {2021})}\BibitemShut {NoStop}%
\bibitem [{\citenamefont {Bulchandani}\ \emph {et~al.}(2021)\citenamefont {Bulchandani}, \citenamefont {Gopalakrishnan},\ and\ \citenamefont {Ilievski}}]{Bulchandani2021}%
  \BibitemOpen
  \bibfield  {author} {\bibinfo {author} {\bibfnamefont {V.~B.}\ \bibnamefont {Bulchandani}}, \bibinfo {author} {\bibfnamefont {S.}~\bibnamefont {Gopalakrishnan}}, \ and\ \bibinfo {author} {\bibfnamefont {E.}~\bibnamefont {Ilievski}},\ }\href {\doibase 10.1088/1742-5468/ac12c7} {\bibfield  {journal} {\bibinfo  {journal} {Journal of Statistical Mechanics: Theory and Experiment}\ }\textbf {\bibinfo {volume} {2021}},\ \bibinfo {pages} {084001} (\bibinfo {year} {2021})}\BibitemShut {NoStop}%
\bibitem [{\citenamefont {Gopalakrishnan}\ and\ \citenamefont {Vasseur}(2023{\natexlab{a}})}]{Gopalakrishnan_2023}%
  \BibitemOpen
  \bibfield  {author} {\bibinfo {author} {\bibfnamefont {S.}~\bibnamefont {Gopalakrishnan}}\ and\ \bibinfo {author} {\bibfnamefont {R.}~\bibnamefont {Vasseur}},\ }\href {\doibase 10.1088/1361-6633/acb36e} {\bibfield  {journal} {\bibinfo  {journal} {Reports on Progress in Physics}\ }\textbf {\bibinfo {volume} {86}},\ \bibinfo {pages} {036502} (\bibinfo {year} {2023}{\natexlab{a}})}\BibitemShut {NoStop}%
\bibitem [{\citenamefont {Gopalakrishnan}\ and\ \citenamefont {Vasseur}(2023{\natexlab{b}})}]{gopalakrishnan2023superdiffusion}%
  \BibitemOpen
  \bibfield  {author} {\bibinfo {author} {\bibfnamefont {S.}~\bibnamefont {Gopalakrishnan}}\ and\ \bibinfo {author} {\bibfnamefont {R.}~\bibnamefont {Vasseur}},\ }\href@noop {} {\enquote {\bibinfo {title} {Superdiffusion from nonabelian symmetries in nearly integrable systems},}\ } (\bibinfo {year} {2023}{\natexlab{b}}),\ \Eprint {http://arxiv.org/abs/2305.15463} {arXiv:2305.15463 [cond-mat.stat-mech]} \BibitemShut {NoStop}%
\bibitem [{\citenamefont {Bertini}\ \emph {et~al.}(2016)\citenamefont {Bertini}, \citenamefont {Collura}, \citenamefont {De~Nardis},\ and\ \citenamefont {Fagotti}}]{Fagotti}%
  \BibitemOpen
  \bibfield  {author} {\bibinfo {author} {\bibfnamefont {B.}~\bibnamefont {Bertini}}, \bibinfo {author} {\bibfnamefont {M.}~\bibnamefont {Collura}}, \bibinfo {author} {\bibfnamefont {J.}~\bibnamefont {De~Nardis}}, \ and\ \bibinfo {author} {\bibfnamefont {M.}~\bibnamefont {Fagotti}},\ }\href {\doibase 10.1103/PhysRevLett.117.207201} {\bibfield  {journal} {\bibinfo  {journal} {Phys. Rev. Lett.}\ }\textbf {\bibinfo {volume} {117}},\ \bibinfo {pages} {207201} (\bibinfo {year} {2016})}\BibitemShut {NoStop}%
\bibitem [{\citenamefont {Castro-Alvaredo}\ \emph {et~al.}(2016)\citenamefont {Castro-Alvaredo}, \citenamefont {Doyon},\ and\ \citenamefont {Yoshimura}}]{Doyon}%
  \BibitemOpen
  \bibfield  {author} {\bibinfo {author} {\bibfnamefont {O.~A.}\ \bibnamefont {Castro-Alvaredo}}, \bibinfo {author} {\bibfnamefont {B.}~\bibnamefont {Doyon}}, \ and\ \bibinfo {author} {\bibfnamefont {T.}~\bibnamefont {Yoshimura}},\ }\href {\doibase 10.1103/PhysRevX.6.041065} {\bibfield  {journal} {\bibinfo  {journal} {Phys. Rev. X}\ }\textbf {\bibinfo {volume} {6}},\ \bibinfo {pages} {041065} (\bibinfo {year} {2016})}\BibitemShut {NoStop}%
\bibitem [{\citenamefont {Doyon}(2020)}]{Doyon_2020}%
  \BibitemOpen
  \bibfield  {author} {\bibinfo {author} {\bibfnamefont {B.}~\bibnamefont {Doyon}},\ }\href {\doibase 10.21468/scipostphyslectnotes.18} {\bibfield  {journal} {\bibinfo  {journal} {SciPost Physics Lecture Notes}\ } (\bibinfo {year} {2020}),\ 10.21468/scipostphyslectnotes.18}\BibitemShut {NoStop}%
\bibitem [{\citenamefont {Bastianello}\ \emph {et~al.}(2022)\citenamefont {Bastianello}, \citenamefont {Bertini}, \citenamefont {Doyon},\ and\ \citenamefont {Vasseur}}]{Bastianello_2022}%
  \BibitemOpen
  \bibfield  {author} {\bibinfo {author} {\bibfnamefont {A.}~\bibnamefont {Bastianello}}, \bibinfo {author} {\bibfnamefont {B.}~\bibnamefont {Bertini}}, \bibinfo {author} {\bibfnamefont {B.}~\bibnamefont {Doyon}}, \ and\ \bibinfo {author} {\bibfnamefont {R.}~\bibnamefont {Vasseur}},\ }\href {\doibase 10.1088/1742-5468/ac3e6a} {\bibfield  {journal} {\bibinfo  {journal} {Journal of Statistical Mechanics: Theory and Experiment}\ }\textbf {\bibinfo {volume} {2022}},\ \bibinfo {pages} {014001} (\bibinfo {year} {2022})}\BibitemShut {NoStop}%
\bibitem [{\citenamefont {Doyon}\ \emph {et~al.}(2023)\citenamefont {Doyon}, \citenamefont {Gopalakrishnan}, \citenamefont {Møller}, \citenamefont {Schmiedmayer},\ and\ \citenamefont {Vasseur}}]{doyon2023generalized}%
  \BibitemOpen
  \bibfield  {author} {\bibinfo {author} {\bibfnamefont {B.}~\bibnamefont {Doyon}}, \bibinfo {author} {\bibfnamefont {S.}~\bibnamefont {Gopalakrishnan}}, \bibinfo {author} {\bibfnamefont {F.}~\bibnamefont {Møller}}, \bibinfo {author} {\bibfnamefont {J.}~\bibnamefont {Schmiedmayer}}, \ and\ \bibinfo {author} {\bibfnamefont {R.}~\bibnamefont {Vasseur}},\ }\href@noop {} {\enquote {\bibinfo {title} {Generalized hydrodynamics: a perspective},}\ } (\bibinfo {year} {2023}),\ \Eprint {http://arxiv.org/abs/2311.03438} {arXiv:2311.03438 [cond-mat.stat-mech]} \BibitemShut {NoStop}%
\bibitem [{\citenamefont {Bulchandani}(2020)}]{vir2019}%
  \BibitemOpen
  \bibfield  {author} {\bibinfo {author} {\bibfnamefont {V.~B.}\ \bibnamefont {Bulchandani}},\ }\href {\doibase 10.1103/PhysRevB.101.041411} {\bibfield  {journal} {\bibinfo  {journal} {Phys. Rev. B}\ }\textbf {\bibinfo {volume} {101}},\ \bibinfo {pages} {041411} (\bibinfo {year} {2020})}\BibitemShut {NoStop}%
\bibitem [{\citenamefont {De~Nardis}\ \emph {et~al.}(2020{\natexlab{a}})\citenamefont {De~Nardis}, \citenamefont {Gopalakrishnan}, \citenamefont {Ilievski},\ and\ \citenamefont {Vasseur}}]{PhysRevLett.125.070601}%
  \BibitemOpen
  \bibfield  {author} {\bibinfo {author} {\bibfnamefont {J.}~\bibnamefont {De~Nardis}}, \bibinfo {author} {\bibfnamefont {S.}~\bibnamefont {Gopalakrishnan}}, \bibinfo {author} {\bibfnamefont {E.}~\bibnamefont {Ilievski}}, \ and\ \bibinfo {author} {\bibfnamefont {R.}~\bibnamefont {Vasseur}},\ }\href {\doibase 10.1103/PhysRevLett.125.070601} {\bibfield  {journal} {\bibinfo  {journal} {Phys. Rev. Lett.}\ }\textbf {\bibinfo {volume} {125}},\ \bibinfo {pages} {070601} (\bibinfo {year} {2020}{\natexlab{a}})}\BibitemShut {NoStop}%
\bibitem [{\citenamefont {Ilievski}\ \emph {et~al.}(2021)\citenamefont {Ilievski}, \citenamefont {De~Nardis}, \citenamefont {Gopalakrishnan}, \citenamefont {Vasseur},\ and\ \citenamefont {Ware}}]{PhysRevX.11.031023}%
  \BibitemOpen
  \bibfield  {author} {\bibinfo {author} {\bibfnamefont {E.}~\bibnamefont {Ilievski}}, \bibinfo {author} {\bibfnamefont {J.}~\bibnamefont {De~Nardis}}, \bibinfo {author} {\bibfnamefont {S.}~\bibnamefont {Gopalakrishnan}}, \bibinfo {author} {\bibfnamefont {R.}~\bibnamefont {Vasseur}}, \ and\ \bibinfo {author} {\bibfnamefont {B.}~\bibnamefont {Ware}},\ }\href {\doibase 10.1103/PhysRevX.11.031023} {\bibfield  {journal} {\bibinfo  {journal} {Phys. Rev. X}\ }\textbf {\bibinfo {volume} {11}},\ \bibinfo {pages} {031023} (\bibinfo {year} {2021})}\BibitemShut {NoStop}%
\bibitem [{\citenamefont {Kardar}\ \emph {et~al.}(1986)\citenamefont {Kardar}, \citenamefont {Parisi},\ and\ \citenamefont {Zhang}}]{kpz}%
  \BibitemOpen
  \bibfield  {author} {\bibinfo {author} {\bibfnamefont {M.}~\bibnamefont {Kardar}}, \bibinfo {author} {\bibfnamefont {G.}~\bibnamefont {Parisi}}, \ and\ \bibinfo {author} {\bibfnamefont {Y.-C.}\ \bibnamefont {Zhang}},\ }\href {\doibase 10.1103/PhysRevLett.56.889} {\bibfield  {journal} {\bibinfo  {journal} {Phys. Rev. Lett.}\ }\textbf {\bibinfo {volume} {56}},\ \bibinfo {pages} {889} (\bibinfo {year} {1986})}\BibitemShut {NoStop}%
\bibitem [{\citenamefont {Ljubotina}\ \emph {et~al.}(2019{\natexlab{b}})\citenamefont {Ljubotina}, \citenamefont {\ifmmode \check{Z}\else \v{Z}\fi{}nidari\ifmmode~\check{c}\else \v{c}\fi{}},\ and\ \citenamefont {Prosen}}]{PhysRevLett.122.210602}%
  \BibitemOpen
  \bibfield  {author} {\bibinfo {author} {\bibfnamefont {M.}~\bibnamefont {Ljubotina}}, \bibinfo {author} {\bibfnamefont {M.}~\bibnamefont {\ifmmode \check{Z}\else \v{Z}\fi{}nidari\ifmmode~\check{c}\else \v{c}\fi{}}}, \ and\ \bibinfo {author} {\bibfnamefont {T.}~\bibnamefont {Prosen}},\ }\href {\doibase 10.1103/PhysRevLett.122.210602} {\bibfield  {journal} {\bibinfo  {journal} {Phys. Rev. Lett.}\ }\textbf {\bibinfo {volume} {122}},\ \bibinfo {pages} {210602} (\bibinfo {year} {2019}{\natexlab{b}})}\BibitemShut {NoStop}%
\bibitem [{\citenamefont {\v{Z}. Krajnik}\ \emph {et~al.}(2020)\citenamefont {\v{Z}. Krajnik}, \citenamefont {Ilievski},\ and\ \citenamefont {Prosen}}]{10.21468/SciPostPhys.9.3.038}%
  \BibitemOpen
  \bibfield  {author} {\bibinfo {author} {\bibnamefont {\v{Z}. Krajnik}}, \bibinfo {author} {\bibfnamefont {E.}~\bibnamefont {Ilievski}}, \ and\ \bibinfo {author} {\bibfnamefont {T.}~\bibnamefont {Prosen}},\ }\href {\doibase 10.21468/SciPostPhys.9.3.038} {\bibfield  {journal} {\bibinfo  {journal} {SciPost Phys.}\ }\textbf {\bibinfo {volume} {9}},\ \bibinfo {pages} {038} (\bibinfo {year} {2020})}\BibitemShut {NoStop}%
\bibitem [{\citenamefont {Fava}\ \emph {et~al.}(2020)\citenamefont {Fava}, \citenamefont {Ware}, \citenamefont {Gopalakrishnan}, \citenamefont {Vasseur},\ and\ \citenamefont {Parameswaran}}]{PhysRevB.102.115121}%
  \BibitemOpen
  \bibfield  {author} {\bibinfo {author} {\bibfnamefont {M.}~\bibnamefont {Fava}}, \bibinfo {author} {\bibfnamefont {B.}~\bibnamefont {Ware}}, \bibinfo {author} {\bibfnamefont {S.}~\bibnamefont {Gopalakrishnan}}, \bibinfo {author} {\bibfnamefont {R.}~\bibnamefont {Vasseur}}, \ and\ \bibinfo {author} {\bibfnamefont {S.~A.}\ \bibnamefont {Parameswaran}},\ }\href {\doibase 10.1103/PhysRevB.102.115121} {\bibfield  {journal} {\bibinfo  {journal} {Phys. Rev. B}\ }\textbf {\bibinfo {volume} {102}},\ \bibinfo {pages} {115121} (\bibinfo {year} {2020})}\BibitemShut {NoStop}%
\bibitem [{\citenamefont {Dupont}\ and\ \citenamefont {Moore}(2020)}]{dupont_moore}%
  \BibitemOpen
  \bibfield  {author} {\bibinfo {author} {\bibfnamefont {M.}~\bibnamefont {Dupont}}\ and\ \bibinfo {author} {\bibfnamefont {J.~E.}\ \bibnamefont {Moore}},\ }\href {\doibase 10.1103/PhysRevB.101.121106} {\bibfield  {journal} {\bibinfo  {journal} {Phys. Rev. B}\ }\textbf {\bibinfo {volume} {101}},\ \bibinfo {pages} {121106} (\bibinfo {year} {2020})}\BibitemShut {NoStop}%
\bibitem [{\citenamefont {Scheie}\ \emph {et~al.}(2021)\citenamefont {Scheie}, \citenamefont {Sherman}, \citenamefont {Dupont}, \citenamefont {Nagler}, \citenamefont {Stone}, \citenamefont {Granroth}, \citenamefont {Moore},\ and\ \citenamefont {Tennant}}]{Scheie2021}%
  \BibitemOpen
  \bibfield  {author} {\bibinfo {author} {\bibfnamefont {A.}~\bibnamefont {Scheie}}, \bibinfo {author} {\bibfnamefont {N.~E.}\ \bibnamefont {Sherman}}, \bibinfo {author} {\bibfnamefont {M.}~\bibnamefont {Dupont}}, \bibinfo {author} {\bibfnamefont {S.~E.}\ \bibnamefont {Nagler}}, \bibinfo {author} {\bibfnamefont {M.~B.}\ \bibnamefont {Stone}}, \bibinfo {author} {\bibfnamefont {G.~E.}\ \bibnamefont {Granroth}}, \bibinfo {author} {\bibfnamefont {J.~E.}\ \bibnamefont {Moore}}, \ and\ \bibinfo {author} {\bibfnamefont {D.~A.}\ \bibnamefont {Tennant}},\ }\href {\doibase 10.1038/s41567-021-01191-6} {\bibfield  {journal} {\bibinfo  {journal} {Nature Physics}\ }\textbf {\bibinfo {volume} {17}},\ \bibinfo {pages} {726} (\bibinfo {year} {2021})}\BibitemShut {NoStop}%
\bibitem [{\citenamefont {Krajnik}\ \emph {et~al.}(2024)\citenamefont {Krajnik}, \citenamefont {Schmidt}, \citenamefont {Ilievski},\ and\ \citenamefont {Prosen}}]{PhysRevLett.132.017101}%
  \BibitemOpen
  \bibfield  {author} {\bibinfo {author} {\bibfnamefont {i.~c.~v.}\ \bibnamefont {Krajnik}}, \bibinfo {author} {\bibfnamefont {J.}~\bibnamefont {Schmidt}}, \bibinfo {author} {\bibfnamefont {E.}~\bibnamefont {Ilievski}}, \ and\ \bibinfo {author} {\bibfnamefont {T.~c.~v.}\ \bibnamefont {Prosen}},\ }\href {\doibase 10.1103/PhysRevLett.132.017101} {\bibfield  {journal} {\bibinfo  {journal} {Phys. Rev. Lett.}\ }\textbf {\bibinfo {volume} {132}},\ \bibinfo {pages} {017101} (\bibinfo {year} {2024})}\BibitemShut {NoStop}%
\bibitem [{\citenamefont {\v{Z}. Krajnik}\ \emph {et~al.}(2022)\citenamefont {\v{Z}. Krajnik}, \citenamefont {Ilievski},\ and\ \citenamefont {Prosen}}]{Krajnik_2022}%
  \BibitemOpen
  \bibfield  {author} {\bibinfo {author} {\bibnamefont {\v{Z}. Krajnik}}, \bibinfo {author} {\bibfnamefont {E.}~\bibnamefont {Ilievski}}, \ and\ \bibinfo {author} {\bibfnamefont {T.~a.}\ \bibnamefont {Prosen}},\ }\href {\doibase 10.1103/physrevlett.128.090604} {\bibfield  {journal} {\bibinfo  {journal} {Physical Review Letters}\ }\textbf {\bibinfo {volume} {128}} (\bibinfo {year} {2022}),\ 10.1103/physrevlett.128.090604}\BibitemShut {NoStop}%
\bibitem [{\citenamefont {De~Nardis}\ \emph {et~al.}(2023)\citenamefont {De~Nardis}, \citenamefont {Gopalakrishnan},\ and\ \citenamefont {Vasseur}}]{PhysRevLett.131.197102}%
  \BibitemOpen
  \bibfield  {author} {\bibinfo {author} {\bibfnamefont {J.}~\bibnamefont {De~Nardis}}, \bibinfo {author} {\bibfnamefont {S.}~\bibnamefont {Gopalakrishnan}}, \ and\ \bibinfo {author} {\bibfnamefont {R.}~\bibnamefont {Vasseur}},\ }\href {\doibase 10.1103/PhysRevLett.131.197102} {\bibfield  {journal} {\bibinfo  {journal} {Phys. Rev. Lett.}\ }\textbf {\bibinfo {volume} {131}},\ \bibinfo {pages} {197102} (\bibinfo {year} {2023})}\BibitemShut {NoStop}%
\bibitem [{\citenamefont {{Google Quantum AI and Collaborators}}(2023)}]{rosenberg2023dynamics}%
  \BibitemOpen
  \bibfield  {author} {\bibinfo {author} {\bibnamefont {{Google Quantum AI and Collaborators}}},\ }\href@noop {} {\enquote {\bibinfo {title} {Dynamics of magnetization at infinite temperature in a heisenberg spin chain},}\ } (\bibinfo {year} {2023}),\ \Eprint {http://arxiv.org/abs/2306.09333} {arXiv:2306.09333 [quant-ph]} \BibitemShut {NoStop}%
\bibitem [{\citenamefont {Friedman}\ \emph {et~al.}(2020)\citenamefont {Friedman}, \citenamefont {Gopalakrishnan},\ and\ \citenamefont {Vasseur}}]{friedman2019diffusive}%
  \BibitemOpen
  \bibfield  {author} {\bibinfo {author} {\bibfnamefont {A.~J.}\ \bibnamefont {Friedman}}, \bibinfo {author} {\bibfnamefont {S.}~\bibnamefont {Gopalakrishnan}}, \ and\ \bibinfo {author} {\bibfnamefont {R.}~\bibnamefont {Vasseur}},\ }\href {\doibase 10.1103/PhysRevB.101.180302} {\bibfield  {journal} {\bibinfo  {journal} {Phys. Rev. B}\ }\textbf {\bibinfo {volume} {101}},\ \bibinfo {pages} {180302} (\bibinfo {year} {2020})}\BibitemShut {NoStop}%
\bibitem [{\citenamefont {Durnin}\ \emph {et~al.}(2021)\citenamefont {Durnin}, \citenamefont {Bhaseen},\ and\ \citenamefont {Doyon}}]{PhysRevLett.127.130601}%
  \BibitemOpen
  \bibfield  {author} {\bibinfo {author} {\bibfnamefont {J.}~\bibnamefont {Durnin}}, \bibinfo {author} {\bibfnamefont {M.~J.}\ \bibnamefont {Bhaseen}}, \ and\ \bibinfo {author} {\bibfnamefont {B.}~\bibnamefont {Doyon}},\ }\href {\doibase 10.1103/PhysRevLett.127.130601} {\bibfield  {journal} {\bibinfo  {journal} {Phys. Rev. Lett.}\ }\textbf {\bibinfo {volume} {127}},\ \bibinfo {pages} {130601} (\bibinfo {year} {2021})}\BibitemShut {NoStop}%
\bibitem [{\citenamefont {Bastianello}\ \emph {et~al.}(2020)\citenamefont {Bastianello}, \citenamefont {De~Nardis},\ and\ \citenamefont {De~Luca}}]{bastianello2020generalised}%
  \BibitemOpen
  \bibfield  {author} {\bibinfo {author} {\bibfnamefont {A.}~\bibnamefont {Bastianello}}, \bibinfo {author} {\bibfnamefont {J.}~\bibnamefont {De~Nardis}}, \ and\ \bibinfo {author} {\bibfnamefont {A.}~\bibnamefont {De~Luca}},\ }\href {\doibase 10.1103/PhysRevB.102.161110} {\bibfield  {journal} {\bibinfo  {journal} {Phys. Rev. B}\ }\textbf {\bibinfo {volume} {102}},\ \bibinfo {pages} {161110} (\bibinfo {year} {2020})}\BibitemShut {NoStop}%
\bibitem [{\citenamefont {Bastianello}\ \emph {et~al.}(2021)\citenamefont {Bastianello}, \citenamefont {Luca},\ and\ \citenamefont {Vasseur}}]{Bastianello_2021}%
  \BibitemOpen
  \bibfield  {author} {\bibinfo {author} {\bibfnamefont {A.}~\bibnamefont {Bastianello}}, \bibinfo {author} {\bibfnamefont {A.~D.}\ \bibnamefont {Luca}}, \ and\ \bibinfo {author} {\bibfnamefont {R.}~\bibnamefont {Vasseur}},\ }\href {\doibase 10.1088/1742-5468/ac26b2} {\bibfield  {journal} {\bibinfo  {journal} {Journal of Statistical Mechanics: Theory and Experiment}\ }\textbf {\bibinfo {volume} {2021}},\ \bibinfo {pages} {114003} (\bibinfo {year} {2021})}\BibitemShut {NoStop}%
\bibitem [{\citenamefont {De~Nardis}\ \emph {et~al.}(2021)\citenamefont {De~Nardis}, \citenamefont {Gopalakrishnan}, \citenamefont {Vasseur},\ and\ \citenamefont {Ware}}]{PhysRevLett_stability}%
  \BibitemOpen
  \bibfield  {author} {\bibinfo {author} {\bibfnamefont {J.}~\bibnamefont {De~Nardis}}, \bibinfo {author} {\bibfnamefont {S.}~\bibnamefont {Gopalakrishnan}}, \bibinfo {author} {\bibfnamefont {R.}~\bibnamefont {Vasseur}}, \ and\ \bibinfo {author} {\bibfnamefont {B.}~\bibnamefont {Ware}},\ }\href {\doibase 10.1103/PhysRevLett.127.057201} {\bibfield  {journal} {\bibinfo  {journal} {Phys. Rev. Lett.}\ }\textbf {\bibinfo {volume} {127}},\ \bibinfo {pages} {057201} (\bibinfo {year} {2021})}\BibitemShut {NoStop}%
\bibitem [{\citenamefont {Roy}\ \emph {et~al.}(2023{\natexlab{a}})\citenamefont {Roy}, \citenamefont {Dhar}, \citenamefont {Spohn},\ and\ \citenamefont {Kulkarni}}]{roy2023nonequilibrium}%
  \BibitemOpen
  \bibfield  {author} {\bibinfo {author} {\bibfnamefont {D.}~\bibnamefont {Roy}}, \bibinfo {author} {\bibfnamefont {A.}~\bibnamefont {Dhar}}, \bibinfo {author} {\bibfnamefont {H.}~\bibnamefont {Spohn}}, \ and\ \bibinfo {author} {\bibfnamefont {M.}~\bibnamefont {Kulkarni}},\ }\href@noop {} {\enquote {\bibinfo {title} {Nonequilibrium spin transport in integrable and non-integrable classical spin chains},}\ } (\bibinfo {year} {2023}{\natexlab{a}}),\ \Eprint {http://arxiv.org/abs/2306.07864} {arXiv:2306.07864 [cond-mat.stat-mech]} \BibitemShut {NoStop}%
\bibitem [{\citenamefont {Roy}\ \emph {et~al.}(2023{\natexlab{b}})\citenamefont {Roy}, \citenamefont {Dhar}, \citenamefont {Spohn},\ and\ \citenamefont {Kulkarni}}]{PhysRevB.107.L100413}%
  \BibitemOpen
  \bibfield  {author} {\bibinfo {author} {\bibfnamefont {D.}~\bibnamefont {Roy}}, \bibinfo {author} {\bibfnamefont {A.}~\bibnamefont {Dhar}}, \bibinfo {author} {\bibfnamefont {H.}~\bibnamefont {Spohn}}, \ and\ \bibinfo {author} {\bibfnamefont {M.}~\bibnamefont {Kulkarni}},\ }\href {\doibase 10.1103/PhysRevB.107.L100413} {\bibfield  {journal} {\bibinfo  {journal} {Phys. Rev. B}\ }\textbf {\bibinfo {volume} {107}},\ \bibinfo {pages} {L100413} (\bibinfo {year} {2023}{\natexlab{b}})}\BibitemShut {NoStop}%
\bibitem [{\citenamefont {McRoberts}\ \emph {et~al.}(2022{\natexlab{a}})\citenamefont {McRoberts}, \citenamefont {Bilitewski}, \citenamefont {Haque},\ and\ \citenamefont {Moessner}}]{PhysRevB.105.L100403}%
  \BibitemOpen
  \bibfield  {author} {\bibinfo {author} {\bibfnamefont {A.~J.}\ \bibnamefont {McRoberts}}, \bibinfo {author} {\bibfnamefont {T.}~\bibnamefont {Bilitewski}}, \bibinfo {author} {\bibfnamefont {M.}~\bibnamefont {Haque}}, \ and\ \bibinfo {author} {\bibfnamefont {R.}~\bibnamefont {Moessner}},\ }\href {\doibase 10.1103/PhysRevB.105.L100403} {\bibfield  {journal} {\bibinfo  {journal} {Phys. Rev. B}\ }\textbf {\bibinfo {volume} {105}},\ \bibinfo {pages} {L100403} (\bibinfo {year} {2022}{\natexlab{a}})}\BibitemShut {NoStop}%
\bibitem [{\citenamefont {De~Nardis}\ \emph {et~al.}(2020{\natexlab{b}})\citenamefont {De~Nardis}, \citenamefont {Medenjak}, \citenamefont {Karrasch},\ and\ \citenamefont {Ilievski}}]{dmki}%
  \BibitemOpen
  \bibfield  {author} {\bibinfo {author} {\bibfnamefont {J.}~\bibnamefont {De~Nardis}}, \bibinfo {author} {\bibfnamefont {M.}~\bibnamefont {Medenjak}}, \bibinfo {author} {\bibfnamefont {C.}~\bibnamefont {Karrasch}}, \ and\ \bibinfo {author} {\bibfnamefont {E.}~\bibnamefont {Ilievski}},\ }\href {\doibase 10.1103/PhysRevLett.124.210605} {\bibfield  {journal} {\bibinfo  {journal} {Phys. Rev. Lett.}\ }\textbf {\bibinfo {volume} {124}},\ \bibinfo {pages} {210605} (\bibinfo {year} {2020}{\natexlab{b}})}\BibitemShut {NoStop}%
\bibitem [{\citenamefont {Glorioso}\ \emph {et~al.}(2021)\citenamefont {Glorioso}, \citenamefont {Delacrétaz}, \citenamefont {Chen}, \citenamefont {Nandkishore},\ and\ \citenamefont {Lucas}}]{10.21468/SciPostPhys.10.1.015}%
  \BibitemOpen
  \bibfield  {author} {\bibinfo {author} {\bibfnamefont {P.}~\bibnamefont {Glorioso}}, \bibinfo {author} {\bibfnamefont {L.~V.}\ \bibnamefont {Delacrétaz}}, \bibinfo {author} {\bibfnamefont {X.}~\bibnamefont {Chen}}, \bibinfo {author} {\bibfnamefont {R.~M.}\ \bibnamefont {Nandkishore}}, \ and\ \bibinfo {author} {\bibfnamefont {A.}~\bibnamefont {Lucas}},\ }\href {\doibase 10.21468/SciPostPhys.10.1.015} {\bibfield  {journal} {\bibinfo  {journal} {SciPost Phys.}\ }\textbf {\bibinfo {volume} {10}},\ \bibinfo {pages} {015} (\bibinfo {year} {2021})}\BibitemShut {NoStop}%
\bibitem [{\citenamefont {Claeys}\ \emph {et~al.}(2022)\citenamefont {Claeys}, \citenamefont {Lamacraft},\ and\ \citenamefont {Herzog-Arbeitman}}]{PhysRevLett.128.246603}%
  \BibitemOpen
  \bibfield  {author} {\bibinfo {author} {\bibfnamefont {P.~W.}\ \bibnamefont {Claeys}}, \bibinfo {author} {\bibfnamefont {A.}~\bibnamefont {Lamacraft}}, \ and\ \bibinfo {author} {\bibfnamefont {J.}~\bibnamefont {Herzog-Arbeitman}},\ }\href {\doibase 10.1103/PhysRevLett.128.246603} {\bibfield  {journal} {\bibinfo  {journal} {Phys. Rev. Lett.}\ }\textbf {\bibinfo {volume} {128}},\ \bibinfo {pages} {246603} (\bibinfo {year} {2022})}\BibitemShut {NoStop}%
\bibitem [{\citenamefont {Nandy}\ \emph {et~al.}(2023)\citenamefont {Nandy}, \citenamefont {Lenar\ifmmode \check{c}\else \v{c}\fi{}i\ifmmode~\check{c}\else \v{c}\fi{}}, \citenamefont {Ilievski}, \citenamefont {Mierzejewski}, \citenamefont {Herbrych},\ and\ \citenamefont {Prelov\ifmmode~\check{s}\else \v{s}\fi{}ek}}]{PhysRevB.108.L081115}%
  \BibitemOpen
  \bibfield  {author} {\bibinfo {author} {\bibfnamefont {S.}~\bibnamefont {Nandy}}, \bibinfo {author} {\bibfnamefont {Z.}~\bibnamefont {Lenar\ifmmode \check{c}\else \v{c}\fi{}i\ifmmode~\check{c}\else \v{c}\fi{}}}, \bibinfo {author} {\bibfnamefont {E.}~\bibnamefont {Ilievski}}, \bibinfo {author} {\bibfnamefont {M.}~\bibnamefont {Mierzejewski}}, \bibinfo {author} {\bibfnamefont {J.}~\bibnamefont {Herbrych}}, \ and\ \bibinfo {author} {\bibfnamefont {P.}~\bibnamefont {Prelov\ifmmode~\check{s}\else \v{s}\fi{}ek}},\ }\href {\doibase 10.1103/PhysRevB.108.L081115} {\bibfield  {journal} {\bibinfo  {journal} {Phys. Rev. B}\ }\textbf {\bibinfo {volume} {108}},\ \bibinfo {pages} {L081115} (\bibinfo {year} {2023})}\BibitemShut {NoStop}%
\bibitem [{\citenamefont {Ishimori}(1982)}]{Ishimori1982}%
  \BibitemOpen
  \bibfield  {author} {\bibinfo {author} {\bibfnamefont {Y.}~\bibnamefont {Ishimori}},\ }\href {\doibase 10.1143/jpsj.51.3417} {\bibfield  {journal} {\bibinfo  {journal} {Journal of the Physical Society of Japan}\ }\textbf {\bibinfo {volume} {51}},\ \bibinfo {pages} {3417} (\bibinfo {year} {1982})}\BibitemShut {NoStop}%
\bibitem [{\citenamefont {\v{Z}. Krajnik}\ and\ \citenamefont {Prosen}(2020)}]{Krajnik2020rot}%
  \BibitemOpen
  \bibfield  {author} {\bibinfo {author} {\bibnamefont {\v{Z}. Krajnik}}\ and\ \bibinfo {author} {\bibfnamefont {T.}~\bibnamefont {Prosen}},\ }\href {\doibase 10.1007/s10955-020-02523-1} {\bibfield  {journal} {\bibinfo  {journal} {Journal of Statistical Physics}\ }\textbf {\bibinfo {volume} {179}},\ \bibinfo {pages} {110–130} (\bibinfo {year} {2020})}\BibitemShut {NoStop}%
\bibitem [{sup()}]{suppmat}%
  \BibitemOpen
  \href@noop {} {}\bibinfo {note} {See Supplementary Information for further information on discrete time numerics, anisotropic perturbations, additional autocorrelation function data, a discussion of the choice of scaling parameter, and numerics for many-body soliton decay processed for Floquet perturbations.}\BibitemShut {Stop}%
\bibitem [{\citenamefont {De~Nardis}\ \emph {et~al.}(2019)\citenamefont {De~Nardis}, \citenamefont {Medenjak}, \citenamefont {Karrasch},\ and\ \citenamefont {Ilievski}}]{PhysRevLett.123.186601}%
  \BibitemOpen
  \bibfield  {author} {\bibinfo {author} {\bibfnamefont {J.}~\bibnamefont {De~Nardis}}, \bibinfo {author} {\bibfnamefont {M.}~\bibnamefont {Medenjak}}, \bibinfo {author} {\bibfnamefont {C.}~\bibnamefont {Karrasch}}, \ and\ \bibinfo {author} {\bibfnamefont {E.}~\bibnamefont {Ilievski}},\ }\href {\doibase 10.1103/PhysRevLett.123.186601} {\bibfield  {journal} {\bibinfo  {journal} {Phys. Rev. Lett.}\ }\textbf {\bibinfo {volume} {123}},\ \bibinfo {pages} {186601} (\bibinfo {year} {2019})}\BibitemShut {NoStop}%
\bibitem [{\citenamefont {Ilievski}\ \emph {et~al.}(2018)\citenamefont {Ilievski}, \citenamefont {De~Nardis}, \citenamefont {Medenjak},\ and\ \citenamefont {Prosen}}]{idmp}%
  \BibitemOpen
  \bibfield  {author} {\bibinfo {author} {\bibfnamefont {E.}~\bibnamefont {Ilievski}}, \bibinfo {author} {\bibfnamefont {J.}~\bibnamefont {De~Nardis}}, \bibinfo {author} {\bibfnamefont {M.}~\bibnamefont {Medenjak}}, \ and\ \bibinfo {author} {\bibfnamefont {T.}~\bibnamefont {Prosen}},\ }\href {\doibase 10.1103/PhysRevLett.121.230602} {\bibfield  {journal} {\bibinfo  {journal} {Phys. Rev. Lett.}\ }\textbf {\bibinfo {volume} {121}},\ \bibinfo {pages} {230602} (\bibinfo {year} {2018})}\BibitemShut {NoStop}%
\bibitem [{\citenamefont {Sachdev}\ and\ \citenamefont {Damle}(1997)}]{PhysRevLett.78.943}%
  \BibitemOpen
  \bibfield  {author} {\bibinfo {author} {\bibfnamefont {S.}~\bibnamefont {Sachdev}}\ and\ \bibinfo {author} {\bibfnamefont {K.}~\bibnamefont {Damle}},\ }\href {\doibase 10.1103/PhysRevLett.78.943} {\bibfield  {journal} {\bibinfo  {journal} {Phys. Rev. Lett.}\ }\textbf {\bibinfo {volume} {78}},\ \bibinfo {pages} {943} (\bibinfo {year} {1997})}\BibitemShut {NoStop}%
\bibitem [{\citenamefont {Damle}\ and\ \citenamefont {Sachdev}(1998)}]{PhysRevB.57.8307}%
  \BibitemOpen
  \bibfield  {author} {\bibinfo {author} {\bibfnamefont {K.}~\bibnamefont {Damle}}\ and\ \bibinfo {author} {\bibfnamefont {S.}~\bibnamefont {Sachdev}},\ }\href {\doibase 10.1103/PhysRevB.57.8307} {\bibfield  {journal} {\bibinfo  {journal} {Phys. Rev. B}\ }\textbf {\bibinfo {volume} {57}},\ \bibinfo {pages} {8307} (\bibinfo {year} {1998})}\BibitemShut {NoStop}%
\bibitem [{\citenamefont {Agrawal}\ \emph {et~al.}(2020)\citenamefont {Agrawal}, \citenamefont {Gopalakrishnan}, \citenamefont {Vasseur},\ and\ \citenamefont {Ware}}]{PhysRevB.101.224415}%
  \BibitemOpen
  \bibfield  {author} {\bibinfo {author} {\bibfnamefont {U.}~\bibnamefont {Agrawal}}, \bibinfo {author} {\bibfnamefont {S.}~\bibnamefont {Gopalakrishnan}}, \bibinfo {author} {\bibfnamefont {R.}~\bibnamefont {Vasseur}}, \ and\ \bibinfo {author} {\bibfnamefont {B.}~\bibnamefont {Ware}},\ }\href {\doibase 10.1103/PhysRevB.101.224415} {\bibfield  {journal} {\bibinfo  {journal} {Phys. Rev. B}\ }\textbf {\bibinfo {volume} {101}},\ \bibinfo {pages} {224415} (\bibinfo {year} {2020})}\BibitemShut {NoStop}%
\bibitem [{\citenamefont {Lakshmanan}\ \emph {et~al.}(1976)\citenamefont {Lakshmanan}, \citenamefont {Ruijgrok},\ and\ \citenamefont {Thompson}}]{lrt}%
  \BibitemOpen
  \bibfield  {author} {\bibinfo {author} {\bibfnamefont {M.}~\bibnamefont {Lakshmanan}}, \bibinfo {author} {\bibfnamefont {T.~W.}\ \bibnamefont {Ruijgrok}}, \ and\ \bibinfo {author} {\bibfnamefont {C.}~\bibnamefont {Thompson}},\ }\href@noop {} {\bibfield  {journal} {\bibinfo  {journal} {Physica A: Statistical Mechanics and its Applications}\ }\textbf {\bibinfo {volume} {84}},\ \bibinfo {pages} {577} (\bibinfo {year} {1976})}\BibitemShut {NoStop}%
\bibitem [{\citenamefont {McRoberts}\ \emph {et~al.}(2022{\natexlab{b}})\citenamefont {McRoberts}, \citenamefont {Bilitewski}, \citenamefont {Haque},\ and\ \citenamefont {Moessner}}]{PhysRevE.106.L062202}%
  \BibitemOpen
  \bibfield  {author} {\bibinfo {author} {\bibfnamefont {A.~J.}\ \bibnamefont {McRoberts}}, \bibinfo {author} {\bibfnamefont {T.}~\bibnamefont {Bilitewski}}, \bibinfo {author} {\bibfnamefont {M.}~\bibnamefont {Haque}}, \ and\ \bibinfo {author} {\bibfnamefont {R.}~\bibnamefont {Moessner}},\ }\href {\doibase 10.1103/PhysRevE.106.L062202} {\bibfield  {journal} {\bibinfo  {journal} {Phys. Rev. E}\ }\textbf {\bibinfo {volume} {106}},\ \bibinfo {pages} {L062202} (\bibinfo {year} {2022}{\natexlab{b}})}\BibitemShut {NoStop}%
\bibitem [{\citenamefont {McRoberts}\ and\ \citenamefont {Moessner}()}]{adams_draft}%
  \BibitemOpen
  \bibfield  {author} {\bibinfo {author} {\bibfnamefont {A.~J.}\ \bibnamefont {McRoberts}}\ and\ \bibinfo {author} {\bibfnamefont {R.}~\bibnamefont {Moessner}},\ }\href@noop {} {\enquote {\bibinfo {title} {Long lifetime of superdiffusion in non-integrable spin chains},}\ }\BibitemShut {NoStop}%
\end{thebibliography}%

\bigskip

\onecolumngrid
\newpage



\section*{Supplementary material}

# Supplemental material for "Slow crossover from superdiffusion to diffusion in isotropic spin chains"

Catherine McCarthy and Romain Vasseur
*Department of Physics, University of Massachusetts, Amherst, MA 01003, USA*

Sarang Gopalakrishnan
*Department of Electrical and Computer Engineering,
Princeton University, Princeton, NJ 08544, USA*
(Dated: February 23, 2024)

## I. DISCRETE TIME NUMERICS

Here we will give a brief description of the discrete time (Floquet or noisy) model used in this work. This model was recently developed in 1 and 2; more technical details can be found there.

Like other classical integration schemes, each site in the discrete time model hosts an $O(3)$ vector $\vec{S}_n = (S_n^x, S_n^y, S_n^z)$ of unit norm that represents a classical spin. The main object of this model is classical map that time-evolves adjacent pairs of spins; in analogy with quantum circuits, we will refer to this object as the gate $U^{(\text{int})}$. $U^{(\text{int})}$ automatically preserves spin norm, spin-rotational invariance, and integrablity. Acting on a pair of spins $\vec{S}_n$, $\vec{S}_{n+1}$ with $U^{(\text{int})}$ gives

$$U^{(\text{int})}(\vec{S}_n, \vec{S}_{n+1}) = \frac{1}{\sigma^2 + \tau^2}\left((\sigma^2 \vec{S}_n + \tau^2 \vec{S}_{n+1} + \tau \vec{S}_n \times \vec{S}_{n+1}), (\sigma^2 \vec{S}_{n+1} + \tau^2 \vec{S}_n + \tau \vec{S}_{n+1} \times \vec{S}_n)\right), \tag{S1}$$

where $\sigma^2$ is defined by $\sigma^2 = \frac{1}{2}(1 + \vec{S}_n \cdot \vec{S}_{n+1})$.

The parameter $\tau$ can be understood as a discrete timestep. Successively applying two layers of $U^{(\text{int})}$ gates to neighboring spins in a brickwork (odd-even) pattern corresponds to updating the system by one timestep of duration $\tau$ (Fig. S1). Taking the limit $\tau \to 0$ recovers continuous time evolution under the integrable Ishimori Hamiltonian[1]. Note that in this model, any finite value of $\tau$ preserves integrability, even if energy is not conserved. Throughout this work, we use $\tau = 0.25$, which is small enough to produce $z(t)$ curves that strongly overlap in both the integrable and nonintegrable limits with smaller values of $\tau$.

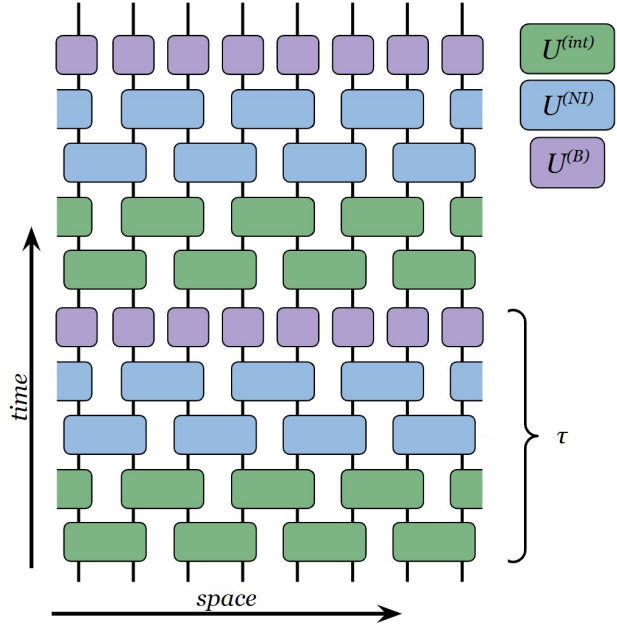

FIG. S1: **Discrete time model.** Brickwork pattern of classical gates defining the classical model studied in this work.

We break integrability by adding nonintegrable classical gates. For the SO(3)-respecting ("isotropic") nonintegrable perturbations studied in the main text, we introduce the two-site gate $U^{(\text{NI})}(\tau_m)$ at time $\tau_m = m\tau$ acting on sites $n$ and $n+1$, that corresponds to solving the equations of motion of the (nonintegrable) classical Heisenberg interaction $V_{\text{NI}} = \sum_n g_n(\tau_m) \vec{S}_n \cdot \vec{S}_{n+1}$ for time $t = \tau$[3]. Here $g_n(\tau_m)$ corresponds to either a random parameter (in space and in time $\tau_m = m\tau$ chosen from a normal distribution of unit width centered around 1: noisy model) or is set to one (Floquet model). $U^{(\text{NI})}$ is applied in the same brickwork pattern as $U^{(\text{int})}$ at each time step to perform an update $\tau_m \to \tau_{m+1} = \tau_m + \tau$ (Fig. S1).

We will also use anisotropic (i.e. SO(3)-breaking) perturbations that can be implemented in this model as a single layer of single-site gates $U^{(\text{B})}$ that apply a staggered field of uniform strength $b$ along direction $\hat{e}_z$ to each site in the system. This is analogous to adding the term $V_{\text{anisotropic}} = b \sum_n (-1)^n \vec{S}_n \cdot \hat{e}_z$ to the system's Hamiltonian, which conserves $S^z_{\text{tot}}$ while breaking conservation of $S^x_{\text{tot}}$ and $S^y_{\text{tot}}$. The single-site gates $U_B$ are obtained from integrating the Hamiltonian $(-1)^n \vec{S}_n \cdot \hat{e}_z$ up to time $t = \tau$.

The numerics in this work are implemented by successively layering $U^{(\text{int})}$, $U^{(\text{NI})}$, and $U^{(\text{B})}$, as shown in Figure S1. The strength of $U^{(\text{NI})}$ and $U^{(\text{B})}$ is controlled by parameters $\lambda$ and $b$ respectively. In the present work, we consider either an isotropic or an anisotropic perturbation, but never both at the same time (i.e. we take either $\lambda = 0$ or $b=0$). The perturbation strength enters the model through adjusting the amount of time for which each species of gate is applied. More precisely, when applying an isotropic perturbation of strength $\lambda$, each timestep $\tau$ is decomposed into $\tau = \tau^{(\text{int})} + \tau^{(\text{NI})}$, where $\tau^{(\text{int})} = (1-\lambda)\tau$ is the amount of time that the integrable gate $U^{(\text{int})}$ acts and $\tau^{(\text{NI})} = \lambda\tau$ is the timestep associated with the nonintegrable gate $U^{(\text{NI})}$.

## II. ANISOTROPIC PERTURBATIONS

We expect anisotropic perturbations of strength $b$ to induce a crossover to diffusive transport characterized by a timescale $t_\star \sim b^{-2}$, according to standard Fermi's Golden Rule (FGR) arguments in quantum systems[4]. To confirm that these predictions hold for long times also in our classical model, we implement the anisotropic perturbation described in the previous section. We observe a clear collapse for the effective dynamical exponent $z(t)$ (extracted from the variance of spin transfer, as explained in the main text) when rescaled against $t/t_\star \sim tb^2$ (fig. S2).

As discussed in the main text, the crossover timescale for an anisotropic perturbation of strength $b$ can be understood as a competition between the integrable screening rate $\Gamma^0_s \sim s^{-3}$ for a soliton of size $s$ and the perturbative decay rate $\Gamma_b$. Our data is compatible with the a decay rate independent of the size $s$, and $\Gamma_b \sim b^2$[4]. Since solitons of size $s$ are fully screened and cease to contribute to spin transport at $t_\star \sim s^3$, by matching the screening and perturbative terms $s^{-3} \sim b^2$, we obtain the observed scaling relation $t_\star \sim b^{-2}$.

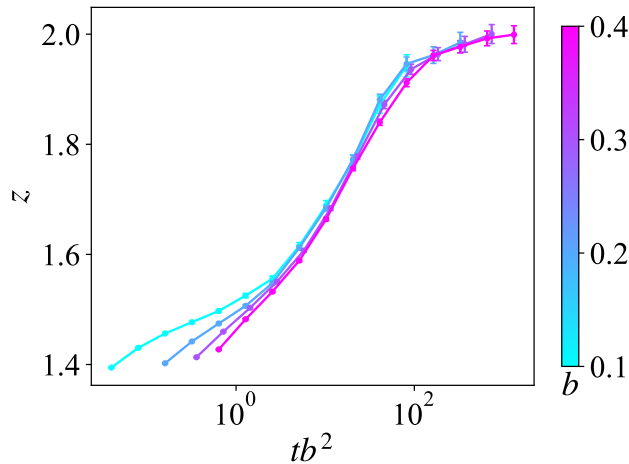

FIG. S2: **Crossover to diffusion for anisotropic perturbations.** Evolution of dynamical exponent $z(t)$ under SO(3)-breaking perturbations of strength $b$. Simulations were run for a system size of N=25000 up to a total time of $t_{\max} = 2^{13}$, and data is averaged over 1000 random initial states. Dashed horizontal lines correspond to superdiffusive $z = 3/2$ and diffusive $z = 2$ transport. Rescaling by $t \cdot b^2$ yields a curve collapse, indicating a crossover timescale $t_\star \sim b^{-2}$.

## III. ADDITIONAL AUTOCORRELATION FUNCTION DATA

In the main text, we primarily computed the dynamical exponent $z(t)$ from fluctuations of spin transfer across the links between adjacent sites in our system. The dynamical exponent may also be computed from other quantities, including the autocorrelation function, with is given by $C(t) = \frac{1}{N} \sum_n \langle S_n^z(t) S_n^z(0) \rangle$ for a system with $N$ sites; it is expected to scale as $C(t) \sim t^{-1/z}$. We find that extracting an effective time-dependent dynamical exponent $z(t)$ from moving fits (or from logarithmic derivatives) of the autocorrelation function requires significantly better statistics than computing $z(t)$ from the fluctuation of spin transfer. Despite poor data resolution at long times, extracting $z(t)$ from the autocorrelation function $C(t)$ yields a crossover compatible with $t_\star \sim \lambda^{-6}$ scaling (figs. S3 and S4) for both noisy and Floquet isotropic perturbations, in agreement with the numerics featured in the main text.

For noisy isotropic perturbations, we also present data for the extracted diffusion constant $D(\lambda)$ (Fig. S4) – see the main text for the Floquet data. As in the main text, we note that at long times $t \gg t_\star$, the autocorrelation function approaches the diffusive form $C(t) \sim 1/\sqrt{D(\lambda)t} + \beta/t + \ldots$ when absorbing irrelevant prefactors into the diffusion constant $D(\lambda)$ and where the $1/t$ corrections arise from standard hydrodynamic tails arguments[5]. For a crossover time scale $t_\star \sim \lambda^{-\alpha}$, by matching the diffusive and superdiffusive behavior at $t = t_\star$, we see that the diffusion constant should scale with $\lambda$ as $D(\lambda) \sim t_\star^{1/3} \sim \lambda^{-\alpha/3}$. Extracting $1/\sqrt{D(\lambda)}$ from long-time fits of $C(t)$, we find that $1/\sqrt{D(\lambda)} \sim \lambda^{\alpha/6}$ is compatible with a linear scaling with $\lambda$ as $\lambda \to 0$, consistent with our other results indicating that $\alpha \approx 6$.

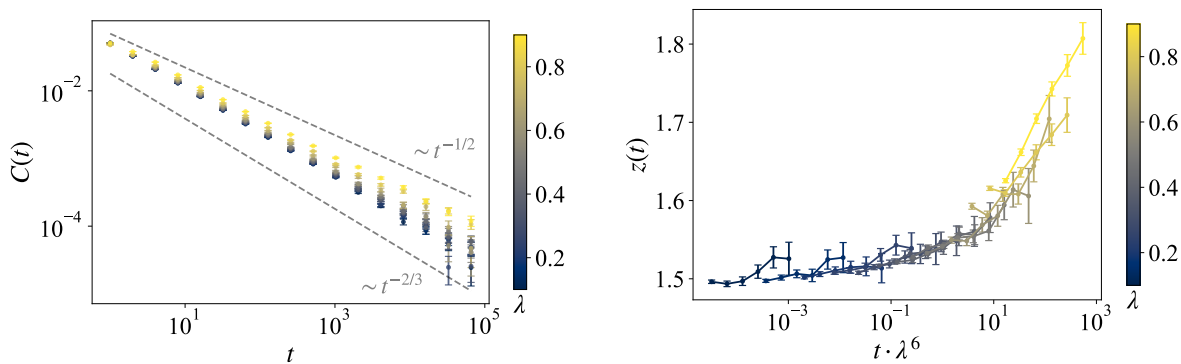

FIG. S3: **Autocorrelation function for Floquet (non-noisy) isotropic perturbations.** Left: Autocorrelation function $C(t) = \frac{1}{N} \sum_n^N \langle S_n^z(t) S_n^z(0) \rangle$ for Floquet isotropic perturbations of strength $\lambda = 0.1$ to $\lambda = 0.9$. Here, the system size is $N = 200000$, and the numerics are averaged over $\sim 4000$ samples. Right: Effective time-dependent dynamical exponent $z(t)$ extracted from the autocorrelation function $C(t)$ from a moving fit for times up to $t = 2^{14}$. Despite being a noisier metric, we observe a curve collapse compatible with $t_\star \sim \lambda^{-6}$ scaling, consistent with the collapse discussed in the main text.

## IV. CHOICE OF SCALING PARAMETER

Throughout this work, we perform our curve collapses with the parameter $\lambda$ characterizing the strength of the perturbation, with $H = (1-\lambda)H_0 + \lambda V$ in the Hamiltonian limit. Another natural parametrization of the strength of the perturbation is $g = \frac{\lambda}{1-\lambda}$, so $H/\lambda = H_0 + gV$. In principle, both $g$ or $\lambda$ are valid choices for scaling parameters as $g \sim \lambda$ in the scaling regime $\lambda \to 0$. In practice though, some values of $\lambda$ are large enough that using $g = \frac{\lambda}{1-\lambda}$ can lead to significantly different collapses away from the regime $\lambda \to 0$. When using $\frac{\lambda}{1-\lambda}$ as our collapse parameter, we still see collapses that are largely consistent with $t_\star \sim \lambda^{-\alpha}$ with $\alpha = 6$ scaling in the Floquet case (Fig. S5, left), although the collapse is significantly worse for larger values of $\lambda$. Smaller values of $\alpha$ (such as $\alpha = 4$ or even $\alpha = 3$) lead to good collapses for larger values of $\lambda$, but fail to collapse curves in the relevant scaling regime $t \to \infty$ and $\lambda \to 0$ with $t\lambda^\alpha$ fixed.

In the noisy case, the collapse is significantly worse using $g = \frac{\lambda}{1-\lambda}$ as a scaling variable (Fig. S5, middle panel), even though the exponent $\alpha = 6$ is the expected answer from perturbative arguments[4]. Another indication that $\lambda$ is a better choice of scaling parameter compared to $g$ is the failure of the collapse of the vacuum soliton decay rates $\Gamma/g^2 \sim \Gamma/\lambda^2$ (Fig. S5, right panel). Given the solid theoretical basis for expecting the vacuum decay rates to scale

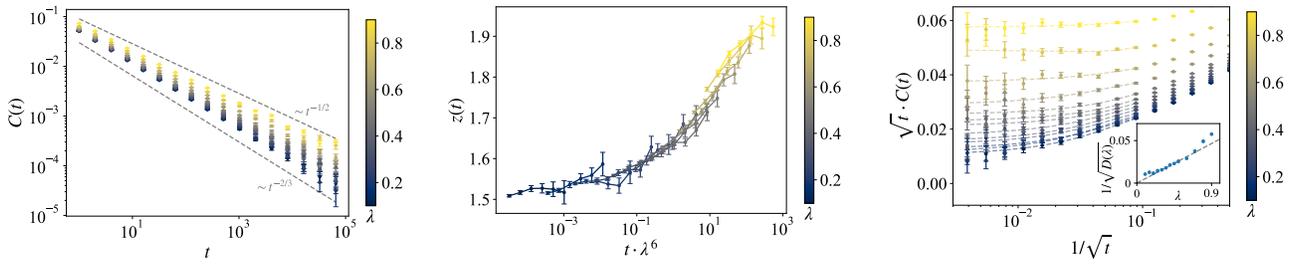

FIG. S4: **Autocorrelation function for noisy perturbations.** Left: Autocorrelation function $C(t) = \frac{1}{N}\sum_n^N \langle S_n^z(t) S_n^z(0)\rangle$ for noisy isotropic perturbations of strength $\lambda = 0.1$ to $\lambda = 0.9$. System size is $N = 200000$, and data is averaged over $\sim 2000$ samples. Middle: Dynamical exponent $z(t)$ extracted from the autocorrelation function $C(t)$ from a moving fit. Similarly to the case of the Floquet perturbation shown in Fig. S3, we observe that $z(t)$ extracted in this way also obeys a curve collapse consistent with the $t_\star \sim \lambda^{-6}$ scaling discussed in the main text. Right: Diffusive long-time limit of the autocorrelation function $C(t)$. Inset: The extracted diffusion constant $D(\lambda)$ is compatible with the scaling $1/\sqrt{D(\lambda)} \sim \lambda^{\alpha/6}$ with $\alpha \approx 6$, in agreement with the crossover time scale $t_\star \sim \lambda^{-6}$.

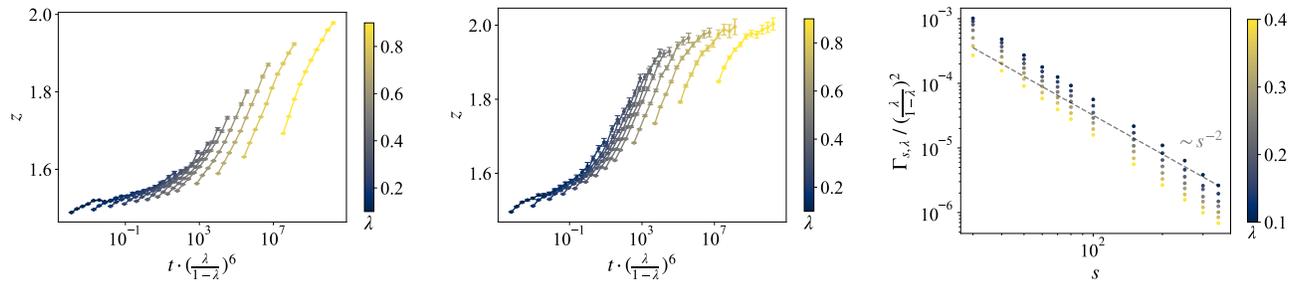

FIG. S5: **Parameter-dependent scaling.** Left: Collapse of the dynamical exponent using $g = \frac{\lambda}{1-\lambda}$ for Floquet isotropic perturbations. Middle: Failed collapse of the dynamical exponent using $g = \frac{\lambda}{1-\lambda}$ for noisy isotropic perturbations. Right: vacuum decay rates $\Gamma_{s,\lambda}$ extracted from the IPR. The decay rates do not collapse when rescaling by $(\lambda/(1-\lambda))^2$, in contrast to when rescaled by $\lambda^2$ (see Fig. 1 in main text). This indicates that $\lambda$ may be a better choice of parameter to use in collapses.

with the square of the perturbation strength[4], the fact that this collapse works well when choosing $\lambda$ and looks worse when choosing $g = \frac{\lambda}{1-\lambda}$ motivates the choice of $\lambda$ as a "good" scaling parameter for numerics with moderately small values of $\lambda$.

## V. ALTERNATIVE CURVE COLLAPSES FOR FLOQUET PERTURBATIONS

In the main text, we observe a crossover timescale that is largely consistent with $t_\star \sim \lambda^{-\alpha}$ with $\alpha = 6$ for both noisy and Floquet perturbations; however, the curve collapses themselves are not able to conclusively rule out other values of $\alpha$ close to $\alpha = 6$. While $\alpha=6$ has a perturbative explanation for noisy perturbations, there is currently no analogous expectation for Floquet perturbations. For completeness, we present additional curve collapses for other values of $\alpha$ close to $\alpha = 6$ for Floquet perturbations in Fig. S6.

## VI. MANY-BODY SOLITON DECAY RATES FOR FLOQUET PERTURBATIONS

Soliton decay rates can be quantified in terms of the inverse participation ratio (IPR) $G(t) = \sum_n (S_n^z(t) - 1)^2$; however, the IPR is only an appropriate metric for single, isolated solitons and loses meaning in the presence of a finite density of solitons. For a noisy perturbation of strength $\lambda$, numerically extracting the vacuum soliton decay rates by the calculating of the IPR is straightforward. In contrast, for Floquet perturbations that do not induce any vacuum decay processes[6], it is more difficult to extract the leading order term of the soliton decay rates, which arise

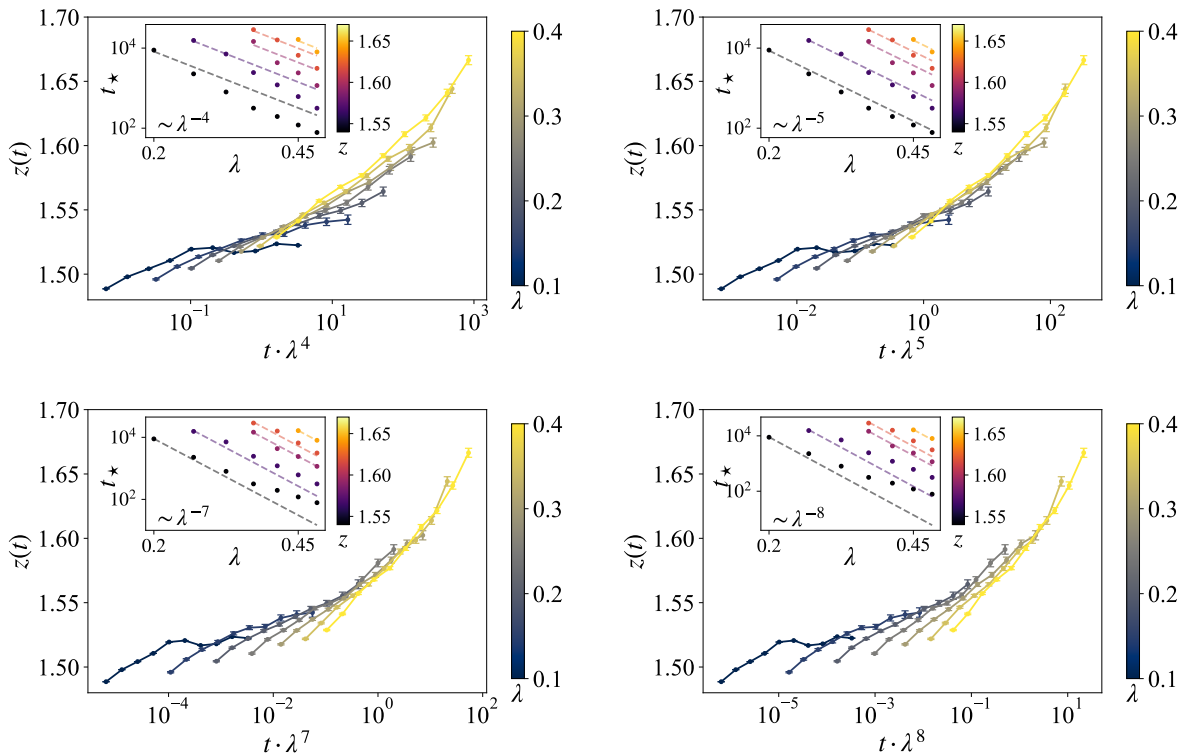

FIG. S6: **Alternative collapses for Floquet perturbation data.** Dynamical exponent $z(t)$ extracted from fluctuations in spin transfer for Floquet perturbations, plotted against $t\lambda^\alpha$ for $\alpha = 4, 5, 7, 8$. The $z(t)$ curve does not collapse for $\alpha = 4$, but the collapses shown for $\alpha = 5, 7$, and 8 look reasonable and may be consistent with $t_\star \sim \lambda^{-\alpha}$ scaling. Insets: timescale $t_\star$ at which a particular value $z(t_\star)$ is reached for different values of $\lambda$, with dashed lines corresponding to power-law fits $t_\star \sim \lambda^{-\alpha}$.

from many-body collisions, using the IPR. An additional complication comes from the decay rates' small magnitude, which makes precisely extracting decay rates from single collisions difficult.

To numerically extract soliton decay rates for Floquet perturbations, we instead study a single soliton in the presence of an infinite-temperature background. The presence of the background, which has an equilibrium density of $\rho_s \sim 1/s^3$ of solitons of size $s$, ensures that any single soliton will experience a large number of collisions and thus measurably decay. Additionally, the leading order term of the decay rate extracted from this setup should be responsible for the crossover timescale for Floquet perturbations discussed in the main text, $t_\star \sim \lambda^{-\alpha}$, with $\alpha \approx 6$. The two principal challenges of this setup are tracking a single soliton and quantifying its decay in an infinite-temperature background, which obscures the individual soliton and prevents the direct calculation of the IPR.

We address these challenges by implementing the following protocol: we initialize a single large soliton of size $s$ in a box of size $N_s$ and place it in the vicinity of $10N_s$ randomly-initialized spins. We then time-evolve the system under a Floquet perturbation of strength $\lambda$ until $t_{\text{stop}}$, at which point we set $\lambda = 0$ and add an additional $N = 11N_s$ vacuum sites to either side of the simulation box. For $\lambda = 0$, the system is integrable and collisions do not lead to decay of either the single large soliton or the solitons that make up the background; therefore, the infinite-temperature background "dissolves" into many small ballistically-propagating solitons. Since solitons move with velocity $v_s \sim 1/s$, the small solitons from the background move faster than the single large soliton. The smaller solitons eventually move into the vacuum region while the single large soliton continues to propagate slowly near the center of the simulation box, which amounts to clearing the background from the region near the single soliton. For numerical tractability, we erase the small solitons that reach the edges of the simulation box and replace them with new vacuum sites. Once the background is cleared from the vicinity of the single large soliton, we calculate the IPR $G(t_{\text{stop}})$. We repeat this procedure for different values of $\lambda$, $s$, and $t_{\text{stop}}$ to construct the IPR for solitons in the presence of an infinite temperature background, which allows the decay rates $\Gamma_{s,\lambda}$ to be extracted from $G(t) \propto \exp(-\Gamma_{s,\lambda} t)$.

We observe a clear $s$-dependence of the extracted decay rates $\Gamma_{s,\lambda}$ (fig. S7), demonstrating that larger solitons retain their longer lifetimes even in thermal states. In lieu of higher-quality data, it is difficult to reliably extract a

scaling exponent, but the decay rates appear to display power-law scaling $\Gamma_{s,\lambda} \sim s^{-\beta}$.

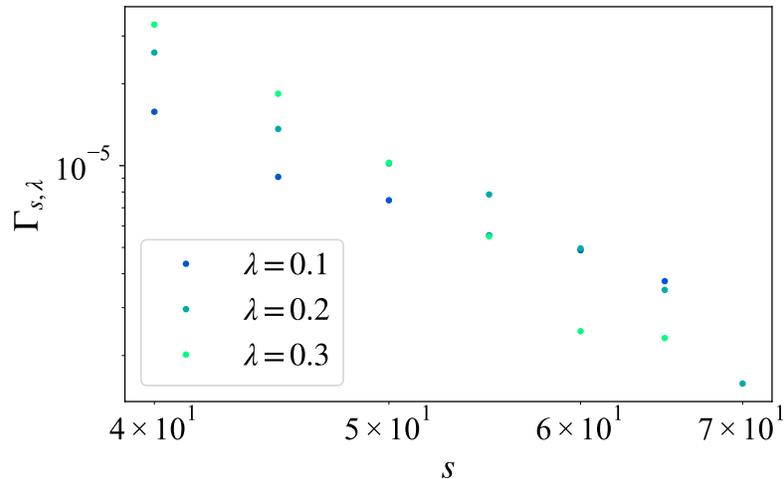

FIG. S7: **Soliton decay rates for Floquet perturbations in the presence of an infinite-temperature background.** Extracted decay rates $\Gamma_{s,\lambda}$ for a soliton of size $s$ in an infinite-temperature background experiencing a perturbation of strength $\lambda$. Each value of $s, \lambda$ is averaged over $\sim 10$ individual realizations. The decay rates appear consistent with power-law scaling $\Gamma_{s,\lambda} \sim \lambda^{-\alpha}$, with larger solitons being longer lived.

---

[1] Ž. Krajnik and T. Prosen, Journal of Statistical Physics **179**, 110–130 (2020).
[2] Ž. Krajnik, E. Ilievski, and T. Prosen, SciPost Phys. **9**, 038 (2020).
[3] Ž. Krajnik, E. Ilievski, and T. a. Prosen, Physical Review Letters **128** (2022), 10.1103/physrevlett.128.090604.
[4] J. De Nardis, S. Gopalakrishnan, R. Vasseur, and B. Ware, Phys. Rev. Lett. **127**, 057201 (2021).
[5] P. Glorioso, L. V. Delacrétaz, X. Chen, R. M. Nandkishore, and A. Lucas, SciPost Phys. **10**, 015 (2021).
[6] A. J. McRoberts, T. Bilitewski, M. Haque, and R. Moessner, Phys. Rev. E **106**, L062202 (2022).



\end{document}